\documentclass[letterpaper,twocolumn,10pt]{article}
\usepackage{usenix-2020-09}

\PassOptionsToPackage{hyphens}{url}\usepackage{hyperref}
\usepackage{xspace}
\usepackage[normalem]{ulem}
\usepackage{adjustbox}
\usepackage[utf8]{inputenc}

\usepackage[scaled=0.85]{beramono} 
\usepackage{amsmath}
\usepackage{xspace}
\usepackage{xcolor}
\usepackage{epsfig}
\usepackage{svg}
\usepackage{subcaption} 
\usepackage{graphicx}
\usepackage{float}
\usepackage{algorithm}
\usepackage{algorithmicx}
\usepackage{algpseudocode}
\usepackage{caption}
\usepackage{xurl}
\usepackage{lipsum}

\usepackage{multirow}
\usepackage{enumitem}
\usepackage{booktabs} 
\usepackage{tabularray}
\usepackage{threeparttable}
\usepackage{cleveref}
\usepackage{tikz}
\usetikzlibrary{automata, positioning, arrows.meta}
\usepackage{verbatim}
\usepackage[T1]{fontenc}

\usepackage[binary-units=true]{siunitx}
\usepackage{listings}
\lstdefinelanguage{gitdiff}{
    morecomment=[f][\color{gray}]{@@},         
    morecomment=[f][\color{red}]{-},           
    morecomment=[f][\color{green!60!black}]{+},
    morecomment=[f][\color{blue}]{diff},      
    morecomment=[f][\color{orange}]{index},   
    morecomment=[f][\color{violet}]{---},     
    morecomment=[f][\color{teal}]{+++},       
}

\usepackage{tcolorbox}
\usepackage{calc}
\crefname{section}{§}{§§}
\crefname{section}{§}{§§}
\crefformat{section}{§#2#1#3}
\crefname{table}{Table}{Table}
\crefname{figure}{Fig.}{Fig.}
\crefname{listing}{Listing}{Listing}
\crefname{algorithm}{Alg.}{Alg.}
\crefname{appendix}{Appendix}{Appendix}
\crefname{objective}{\sc{\textbf{Objective}}}{\sc{\textbf{Objective}}}

\newcommand{\second}[1]{\SI{#1}{\second}}
\newcommand{\byte}[1]{\SI{#1}{\byte}}

\newcommand{\percent}[1]{\SI{#1}{\percent}}

\newcommand{\para}[1]{\vspace{1mm}\noindent\textbf{#1}}

\newcommand{\sysname}{\texttt{Buckler}\xspace}
\newcommand{\name}{\sysname}

\newenvironment{packeditemize}{
\begin{list}{$\bullet$}{
\setlength{\itemsep}{1.5pt}
\setlength{\labelwidth}{8pt}
\setlength{\leftmargin}{10pt}
\setlength{\labelsep}{3pt}
\setlength{\listparindent}{\parindent}
\setlength{\parsep}{1.5pt}
\setlength{\parskip}{1.5pt}
\setlength{\topsep}{1.5pt}}}{\end{list}}

\newtcbox{\customcbox}[1][red]{on line,
arc=3pt,colback=#1!10!white,colframe=#1!50!black,
before upper={\rule[-3pt]{0pt}{10pt}},boxrule=1pt,
boxsep=0pt,left=1pt,right=1pt,top=.2pt,bottom=.2pt}

\definecolor{codegreen}{rgb}{0,0.6,0}
\definecolor{codegray}{rgb}{0.5,0.5,0.5}
\definecolor{codepurple}{rgb}{0.58,0,0.82}
\definecolor{backcolour}{rgb}{0.95,0.95,0.92}
\definecolor{fsmgreen}{HTML}{548235}
\definecolor{fsmred}{HTML}{FF0000}

\lstdefinestyle{mystyle}{
    backgroundcolor=\color{backcolour},   
    commentstyle=\color{codegreen},
    keywordstyle=\color{magenta},
    numberstyle=\tiny\color{codegray},
    stringstyle=\color{codepurple},
    basicstyle=\ttfamily\footnotesize,
    breakatwhitespace=false,         
    breaklines=true,                 
    captionpos=b,                    
    keepspaces=true,                 
    numbers=left,                    
    numbersep=5pt,                  
    showspaces=false,                
    showstringspaces=false,
    showtabs=false,                  
    tabsize=2
}

\DeclareCaptionLabelFormat{newlabel}{#1~#2\ (new)}
\begin{document}

\title{Towards Operator-Empowered Vulnerability Hotfixing \\ for 5G Radio Access Networks}

\author{
{\rm Dong Hyeok Kim}\\
KAIST
\and 
{\rm Xin Zhe Khooi}\\
National University of Singapore
\and 
{\rm Hocheol Nam}\\
KAIST
\and 
{\rm Seungjin Baek}\\
KAIST
\and 
{\rm Mun Choon Chan}\\
National University of Singapore
\and 
{\rm CheolJun Park}\\
Kyung Hee University
\and 
{\rm Min Suk Kang$^*$}\\
KAIST
}

\maketitle

\def\thefootnote{*}\footnotetext{Corresponding author.}

\begin{abstract}
Cellular protocol vulnerabilities can remain exploitable for months or years while standards bodies, vendors, and mobile network operators (MNOs) coordinate permanent fixes.
We present \name, a framework that enables an MNO to deploy temporary, local, and reversible hotfixes in its radio access network (RAN) during this exposure window.
\name places reusable hooks at standardized L2/L3 channel boundaries and exposes a closed, stateful match-action interface with three preventive actions: \textsc{Drop}, \textsc{Modify}, and \textsc{Release}.

We evaluate whether this bounded design provides useful coverage without requiring extensive changes to existing RANs.
From 23 papers, we identify 64 attacks rooted in standard L2/L3 protocol behavior, of which 43 provide a preventive intervention point at the RAN, and we construct \name hotfixes for 20 of them.
All 20 hotfixes use the same rule vocabulary and only five standardized channel hooks, while the unsupported attacks expose endpoint dependencies that a RAN cannot satisfy alone.
We implement the five hooks on srsRAN and OpenAirInterface with small, structurally similar changes, and demonstrate all three actions against representative availability and privacy attacks.
These results establish operator-empowered hotfixing as a practical and portable interim defense and delineate the architectural limits of RAN-only prevention.
\end{abstract}
\section{Introduction}
\label{sec:intro}

Over the last decade, advances in software-defined radios~\cite{Ettus, BladeRF}, open-source cellular stacks~\cite{srsRAN, OAI, OGS}, and accessible experimentation platforms~\cite{2022-WinTECH-POWDER, 2023-comnet-ORANgym} have substantially lowered the barrier to cellular security research.
As a result, the community has discovered a growing number of protocol and implementation vulnerabilities in LTE and 5G systems~\cite{2013-commtut-ltesecsok, 2018-ISSCC-ltevulnsok,2025-WISEC-5gvulnsok}.
Yet, the operational ability to mitigate these vulnerabilities has not improved at the same pace.
When a vulnerability is rooted in standard protocol behavior, a permanent fix requires discussion and agreement within 3GPP, specification changes, vendor adoption, and eventual operator deployment, which can take months or even years~\cite{3GPP-releases}.
Mobile network operators (MNOs) therefore face a systemic \emph{one-day} exposure window in which an attack is already known but not yet preventable in their deployed networks.

In this paper, we argue for an \emph{operator-empowered attack prevention} capability for cellular networks.
MNOs need a practical way to deploy a \emph{temporary} and \emph{local} preventative measure, or a \emph{hotfix}, in their own networks against a known protocol vulnerability.
Such a hotfix would remain active only until a permanent repair becomes available and could be applied to an individual RAN deployment according to its operator's risk assessment.
As with conventional security patches, the hotfix developer and deploying MNO remain responsible for validating that a hotfix prevents the target vulnerable behavior without causing undesirable effects on benign executions, itself a long-studied and non-trivial subject field~\cite{2015-ISSTA-Plausibility, 2012-SE-Dynamictest}. 
We focus on the orthogonal systems problem of constraining, deploying, and withdrawing such an approved rule.
It complements, rather than replaces, standards- and vendor-led remediation by giving the MNO control over its immediate exposure.

Recent advances in Open RAN (or O-RAN) programmability~\cite{oran-arch-site} make this goal more plausible, but do not yet provide the required enforcement capability for attack prevention.
5G-Spector~\cite{2024-NDSS-5GSpector} illustrates both the opportunity and the limitation.
Building on O-RAN's control-plane programmability, 5G-Spector allows an MNO to deploy an xApp that detects previously disclosed L3 attacks using telemetry collected from the RAN.
Its xApp, however, cannot intervene in the RAN message-processing path to prevent the corresponding attack messages from taking effect.
Practical hotfixing requires such timely, inline intervention, but exposing arbitrary access to proprietary RAN internals would be difficult for vendors to support and unsafe for operators to use.
The needed abstraction must, therefore, provide fine-grained message-level enforcement without granting unrestricted programmability.

This paper investigates whether such a design point exists for known L2/L3 cellular protocol attacks.
The central challenge is to provide enough control to prevent a broad range of attacks while requiring only a small, bounded, and portable extension to existing RAN implementations.
We evaluate feasibility along three dimensions:
(1) \emph{Hotfix coverage} asks whether the hooks and bounded rule interface expose enough context and actions to prevent a broad set of RAN-accessible attacks while accommodating newly disclosed ones without new vendor interfaces.
(2) \emph{Deployment footprint} measures the number and size of vendor-side changes and whether the required hooks map portably to standardized boundaries across RAN implementations.
(3) \emph{Operational simplicity} asks whether a closed and predictable rule interface lets an MNO manage hotfixes without arbitrary code execution.

We show that a practical point in this design space exists.
Our framework, called \name\footnote{A buckler is a small handheld shield used to deflect individual blows. The name reflects the proposed framework's role as a bounded defensive tool placed directly in the operator's hands.}, combines reusable hooks at standardized RAN message boundaries with a bounded rule interface that is portable across RAN implementations.
A hotfix developer, either the MNO or an independent party, expresses each hotfix as one or more declarative, stateful match-action rules that match against messages and RAN states, then invoke \textsc{Drop}, \textsc{Modify}, or \textsc{Release}.
The framework's code generator translates this closed vocabulary into inline handlers, so the MNO submits only declarative rules rather than arbitrary executable code. 

We validate \name by systematically attempting to construct hotfixes for attacks in a venue- and year-bounded corpus.
We collect 23 papers reporting 64 distinct attacks rooted in standard L2/L3 protocol behavior; \S\ref{subsec:corpus} details our corpus-selection criteria.
Of these, 43 attacks provide a preventative intervention point at the RAN, and we construct a concrete hotfix for 20 attacks (46.5\%), 
{17 of which still remain unfixed in specification.}
For the other 23 attacks, the most direct countermeasure requires endpoint-dependent authentication or integrity protection (15), confidentiality (6), or verifiable delivery (2), none of which the RAN can introduce alone.
All 20 hotfixes use the same closed rule vocabulary, three actions, and five standardized channel hooks; the hook and action sets remain unchanged for attacks published after 2021.
Applying 5G-Spector's detection methodology to the 43 RAN-intervenable attacks, 25 attacks are preventively detectable at the RAN and 20 of those attacks (80\%) admit a \name hotfix, showing that inline prevention can often complement RAN-side detection. 

We prototype \name on srsRAN and OpenAirInterface (OAI) with an O-RAN SC near-RT RIC, inserting all five channel hooks with 264 and 90 LoC of shared integration code and an average of 12 and 14 additional LoC per hook, respectively.
We experimentally reproduce BTS resource depletion and blind DoS over the air and the CA side channel using RF emulation, exercising all three actions.
Single-hotfix and multi-hotfix measurements show CPU-utilization and RRC-response-latency increases of at most \emph{3.8\%} and \emph{3.37\%}, respectively.

In summary, this paper makes the following contributions:

\begin{itemize}[nosep,leftmargin=*]
\item We formulate operator-empowered hotfixing and design \name, which combines logical hotfix hooks at standardized RAN message boundaries with a bounded, stateful match-action interface.

\item We validate \name against 64 L2/L3 attacks, constructing 20 hotfixes and identifying the endpoint-dependent capabilities that bound RAN-only coverage.

\item We implement \name on two RAN stacks and evaluate its integration footprint, cross-implementation portability, and runtime costs.
\end{itemize}

\section{Motivation and Background}
\label{sec:background}

This section grounds the need and opportunity for operator-empowered hotfixing.
We first explain why MNOs need a local, temporary mitigation during the disclosure-to-deployment interval, then describe how O-RAN makes the 5G RAN programmable yet still lacks the inline enforcement needed for attack prevention.

\subsection{Case for Operator-Empowered Hotfixing}

This work focuses on vulnerabilities rooted in standard protocol behavior.
We exclude implementation vulnerabilities because their causes and remedies are specific to a vendor's codebase and generally require internal implementation knowledge or source-level changes.
Standard protocol vulnerabilities, by contrast, may affect every implementation that follows the vulnerable behavior, giving them potentially ecosystem-wide impact.
Their common message semantics across compliant RAN implementations also make them suitable for the portable hotfixing capability we pursue.

\para{Why MNOs need hotfixing authority.}
Once a standard protocol vulnerability is disclosed, a permanent fix must pass through standardization, release, vendor adoption, and operator deployment, which can outlast responsible disclosure by months or years.
The vulnerability may therefore become public before deployed networks can prevent exploitation.
During this interval, the MNO bears the operational risk but cannot control its exposure.
Operator-empowered hotfixing targets precisely this disclosure-to-deployment exposure window.

\para{Needed by both public and private 5G MNOs.}
Public 5G MNOs can use such a capability to selectively block malicious protocol behavior without affecting legitimate devices.
Private 5G MNOs have an additional need for local control because their security priorities may differ from those of the broader cellular ecosystem.
For example, a vulnerability may pose a critical risk to a military or industrial network yet fail to gain support for a standards-level fix when its ecosystem-wide cost is considered too high.
An operator-empowered hotfix lets such an MNO enforce a local mitigation tailored to its own risk assessment without requiring an immediate change across the entire ecosystem.

\para{A real-world case.}
The carrier aggregation (CA) side-channel attack presented at USENIX Security 2021~\cite{2021-SECURITY-CASideChannel}, for example, illustrates this need.
The attack allows a passive adversary to track a target UE by continuously reading flag bits in downlink MAC CEs.
GSMA confirmed its practicality and impact in 2020 (CVD-2020-0040~\cite{cvd-slic}), yet no fix has been incorporated into the 3GPP specifications.
The vulnerability remained under discussion within 3GPP as recently as November 2025, as shown by several 3GPP S3 Change Requests~\cite{3gpp-s3-254157-keyissue,3gpp-s3-254352-keyissue,3gpp-s3-254446-keyissue}.
This case demonstrates that permanent repair of a protocol vulnerability can remain uncertain for years.
Even if its ecosystem-wide costs outweigh the benefits of a standard fix, a security-sensitive private MNO, such as a military network operator, may reach a different risk assessment and benefit from using \name to mitigate the attack locally.

\subsection{O-RAN Programmability: Opportunities and Limitations}

\para{How O-RAN makes the 5G RAN programmable.}
A 5G network consists of the User Equipment (UE), the Radio Access Network (RAN), which provides wireless connectivity and radio resource control, and the core network, which handles mobility, session control, and service delivery.
Within the RAN protocol stack, the physical layer (L1) handles radio transmission, Layer 2 (L2) comprises MAC, RLC, PDCP, and SDAP, and Layer 3 (L3) comprises RRC and NAS.
The RAN terminates the L2 protocols and RRC, and can therefore observe and act on their messages directly.
NAS terminates at the core network, but some NAS signaling travels transparently inside RRC messages, giving the RAN partial visibility into it.
L1, by contrast, carries physical waveforms rather than standard-defined messages on which an operator can act.
Our RAN-side hotfixing scope is therefore limited to standard L2/L3 protocol behavior.

Within L2 and L3, standard-defined logical channels, including CCCH, DCCH, PCCH, and DTCH, are multiplexed onto transport channels such as UL-SCH and DL-SCH~\cite{3GPP-TS38.321}.
These common message paths provide natural points for observing and acting on protocol messages.
In a conventional 5G RAN, however, message processing is implemented inside vendor-controlled software, with no general interface through which an MNO can program such actions.

Open RAN (or O-RAN) introduces this missing operator-facing programmability.
It restructures the RAN into disaggregated, software-based functions, often implemented as a virtualized RAN (vRAN)~\cite{vran-mavenir,vran-nec,vran-rakuten,vran-samsung}, connected through open, standardized interfaces, and adds the RAN Intelligent Controller (RIC) for running operator-defined control logic~\cite{ORAN-arch}.
The near-real-time RIC hosts operator applications called xApps.
Rather than exposing vendor internals, O-RAN provides predefined controls that xApps access over the standardized E2 interface.
These controls are defined through extensible E2 Service Models (E2SMs)~\cite{ORAN-RC}, allowing an MNO to introduce new control logic without modifying an xApp for each vendor implementation.

\para{Opportunities: L3 attack detection.}
O-RAN's telemetry and xApp abstractions create a promising foundation for operator-developed attack detection.
5G-Spector~\cite{2024-NDSS-5GSpector}, for example, extends the performance telemetry collected through E2SM-KPM~\cite{ORAN-KPM} into a finer, security-focused stream of L3 signaling.
Its shim layer aggregates per-flow suspicious statistics and reports them to an xApp, which detects previously disclosed L3 attacks.
This design demonstrates that O-RAN can expose protocol-level visibility to operator applications at the granularity required for attack detection.

\para{Limitations: Missing in-path enforcement for attack prevention.}
Attack prevention imposes a stronger requirement than detection: the operator must intervene in the RAN execution path to block or sanitize an attack message before it causes damage.
Current O-RAN abstractions primarily expose telemetry and predefined controls, but do not let an xApp enforce general, per-message actions in that path.
Thus, although the 5G-Spector xApp can detect an attack, it cannot stop the corresponding attack messages.

Adding a separate vendor-provided control for every new attack would not provide a practical solution.
A practical hotfix interface must instead be fine-grained enough to mitigate a broad range of vulnerabilities, yet general and bounded to remain compatible across RAN implementations.
This gap motivates \name, which extends O-RAN with standard-message-level enforcement for operator-deployed hotfixes.

\section{Threat Model and Hotfix Requirements}
\label{sec:threat_model}

This section defines the threat model and requirements for practical operator-empowered hotfixing.
We first characterize the adversary's goals and capabilities, and then use the RAN's visibility and control over the relevant protocol messages to establish our scope.
We next elaborate on the three desired properties introduced in~\S\ref{sec:intro}: hotfix coverage, deployment footprint, and operational simplicity.
Finally, we explain how the design space formed by these properties motivates our main contribution.

\subsection{Threat Model}
\label{subsec:threat-model}

\para{Adversary goal and assumptions.}
The adversary's goal is to exploit a known vulnerability in standard L2/L3 cellular protocol behavior to compromise the availability, integrity, or privacy of a legitimate UE or the operator's network.
We assume that the adversary does not compromise the legitimate RAN, core network, the victim UE, or the O-RAN software platform.
Instead, the adversary interacts with the victim UE or the legitimate RAN over the air interface, with capabilities determined by its position in the communication path.

\begin{figure}[t]
    \centerline{\includegraphics[width=0.9\linewidth]{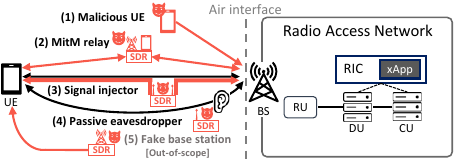}}
    \caption[Overview]{
    Adversary types for L2/L3 cellular protocol attacks.
    Types (1)--(4) expose the relevant protocol messages to the operator's RAN, while (5) does not and is outside our scope.
    }
    \vspace{-12pt}
    \label{fig:threat-model}
\end{figure}

\para{Adversary capabilities.}
Figure~\ref{fig:threat-model} illustrates five adversary types.
Our threat model includes the first four, whose relevant protocol messages reach or traverse the operator's legitimate RAN.
Together, these four types cover the adversary models considered in most cellular RAN attacks reported in the literature, as our corpus analysis later confirms.

\begin{packeditemize}
    \item[(1)] \textbf{Malicious UE.}
    The adversary uses a Commercial Off-the-Shelf (COTS) UE or a software-defined radio (SDR)~\cite{Ettus} to interact with the legitimate RAN as a malicious UE.
    It may launch active attacks using either a legitimate identity registered in the network or a spoofed identity derived by replaying over-the-air messages from a benign UE.

    \item[(2)] \textbf{Man-in-the-Middle (MitM) relay.}
    The adversary deploys a fake base station (FBS) to attract a victim UE and uses a second radio connection to the legitimate RAN.
    Unlike a standalone FBS, this relay forwards messages between the victim UE and the legitimate RAN, allowing the adversary to intercept and potentially modify their communication.

    \item[(3)] \textbf{Signal injector.}
    The adversary uses an SDR to overshadow legitimate RF signals with well-timed, higher-strength transmissions~\cite{2019-USENIX-SigOver,2022-MobiCom-Adaptover}.
    This approach can be more stealthy than a MitM relay while providing similar capabilities to intercept or modify messages exchanged between a benign UE and the RAN.

    \item[(4)] \textbf{Passive eavesdropper.}
    This least-capable adversary uses an SDR or COTS UE to passively monitor over-the-air message exchanges between UEs and the RAN without transmitting signals.
\end{packeditemize}

{
In each of these adversary types, we assume that the adversary does not know and cannot compromise the victim’s cryptographic keys.
}

\para{Scope boundary and non-goals.}
Although their capabilities differ, all four adversary types share the property required by an \emph{RAN-side} hotfix: the attack manifests in standard L2/L3 messages that terminate at or traverse the MNO's legitimate RAN.
As discussed in~\S\ref{sec:background}, the RAN can observe and act on such messages, allowing a hotfix to intervene without modifying the UE or core-network implementation.
This RAN-side vantage point defines the attacks considered in this work.

The fifth adversary type in Figure~\ref{fig:threat-model} uses a standalone fake base station (FBS) and is excluded.
Both types (2) and (5) attract a victim UE to an attacker-controlled FBS, but they differ in whether the victim's messages reach the operator's legitimate RAN.
The MitM relay in type (2) forwards those messages to the RAN, whereas the standalone FBS in type (5) terminates the exchange locally.
Consequently, the RAN can neither observe the exploit messages from type (5) nor enforce a hotfix against them.
For the same reason, we also exclude attacks confined to core-internal protocols, application-layer protocols, or physical-layer signal behavior without standard L2/L3 message semantics, as well as attacks whose mitigation requires changes to UE or core-network implementations.

We also do not aim to secure the O-RAN software platform itself.
Vulnerabilities in the RAN Intelligent Controller (RIC), third-party xApps, O-RAN interfaces, or the software supply chain of programmable RAN components constitute a separate attack surface.
These issues are important but orthogonal to our goal of using O-RAN programmability to hotfix cellular protocol vulnerabilities.
They are addressed by complementary work on O-RAN platform security, including prior studies of RICs, xApps, and O-RAN interfaces; e.g., Yang et al.~\cite{2024-USENIX-ORANalyst}, Xing et al.~\cite{2025-USENIX-fronthaul}, and Atalay et al.~\cite{2023-INFOCOM-xAppAuth}.

\subsection{Desired Properties for Hotfixing}
\label{subsec:desired-properties}

Within this threat model, a practical operator-deployed hotfixing approach should satisfy three properties that capture how many attacks it can prevent, how much implementation effort it requires from RAN vendors, and how tightly its interface bounds the controls exposed to MNOs.

\para{Hotfix coverage.}
The approach should provide sufficient control to prevent a broad range of RAN-accessible L2/L3 protocol attacks.
This requires both \emph{expressiveness} and \emph{extensibility}.
First, its hotfix interface must be expressive enough to encode the preventative measures required by known attacks.
Because these attacks exploit different messages, protocol states, and procedures, the interface must expose the relevant message context and mitigation actions rather than provide only attack-specific, coarse-grained controls.

Second, the hotfixing capability should remain extensible to newly disclosed attacks.
An MNO should be able to express a new hotfix using the existing hooks and supported rule operations, without needing new attack-specific hook or operation.
Otherwise, every newly disclosed attack would require a vendor modification, defeating the purpose of operator-empowered hotfixing.
Coverage should therefore be assessed not only by how many known attacks can be prevented, but also by whether the required hook locations and rule operations remain stable as additional attacks are considered.

\para{Deployment footprint.}
The approach should require little implementation effort from RAN vendors.
Adding hotfixing support to an existing RAN should require only a few hook implementation sites and small source-code changes at each.
Deployment footprint therefore captures both the number of implementation sites and the source-code changes required to realize the logical hooks.

The logical hotfix hooks should also be \emph{portable} across standard-compliant RAN implementations.
Specifically, they should correspond to standardized protocol-message boundaries that every compliant RAN implements, rather than to vendor-specific functions or software structures.
This allows different vendors to realize the same logical hooks even when their internal codebases differ.
Thus, deployment footprint captures not only how much code a vendor must change, but also whether the same small set of logical hooks can be realized across different RAN implementations.

\para{Operational simplicity.}
The approach should strictly bound the degree of freedom exposed through the hotfix interface.
Operational simplicity is not determined merely by the number of hook locations: even a small set would be difficult to validate and operate safely if its rule interface permitted arbitrary code execution or exposed complex, implementation-specific state.
Instead, the rule interface should offer a closed set of clearly defined operations and parameters whose effects on message processing are explicit and auditable.

Such bounded semantics make hotfix rules easier for developers to create and operators to inspect and manage without knowledge of vendor-internal RAN code.
They also allow the control-plane application and the vendor-side RAN implementation to validate rules and reject unsupported operations or malformed parameters before enforcement.
This mechanism-level validation is distinct from establishing that a
particular rule preserves all compliant executions.

\subsection{Our Contribution}

These properties trade off: broader hooks or more permissive rules may improve coverage, but can increase vendor integration effort and make operator controls harder to inspect and manage.
Our contribution is to design and empirically validate \name as a practical point in this space.
Section~\ref{sec:design} derives its standardized hooks and bounded, stateful match-action interface from these requirements, and Section~\ref{sec:attack-corpus-analysis} evaluates the resulting design across the attack corpus by constructing concrete hotfixes where possible and identifying the architectural capability missing otherwise.

\section{\name: A Bounded Hotfix Design}
\label{sec:design}

Section~\ref{sec:threat_model} establishes three properties for practical operator-empowered hotfixing, but satisfying them simultaneously requires careful choices about where hotfixes can intervene and what capabilities they should expose. 
This section presents \name, our bounded design point within this space. 
We first derive its principal design choices from the threat model and desired properties, and then describe its architecture, stateful match-action interface, and mechanisms for deploying and enforcing hotfixes in a running RAN.

\subsection{From Requirements to Design Choices}
\label{sec:rationale}

\name must make two fundamental design decisions: where in the RAN a hotfix
rule may attach and what operations the rule may perform.
We call each permitted attachment point a \emph{hotfix hook}.
A hook is defined logically by a standardized RAN message boundary; a vendor realizes it through a small insertion in the corresponding message-processing path of its vRAN implementation. 
Hook placement therefore determines where a rule can act, while
the bounded rule interface determines what it can observe and do. 
Broadly distributed hooks or unrestricted rule semantics could provide extensive control, but would conflict with the deployment-footprint and operational-simplicity properties. 
\name therefore constrains both aspects as follows.

\para{RAN-bounded, portable hooks at standard message boundaries.}
Given the RAN-side scope established in \S\ref{subsec:threat-model}, \name attaches hotfixes only to standard L2/L3 messages handled by the RAN and uses only protocol state maintained there.
Rather than placing a separate hook in every vulnerable procedure or at vendor-internal functions, \name places hooks at the boundaries of \emph{standardized logical and transport channels}.
Every in-scope message crosses one of these boundaries, and every standard-compliant implementation realizes the corresponding processing path despite differences in internal code structure.
This finite, portable hook set provides broad message visibility while bounding the number and placement of insertion points.

\para{Bounding hotfix semantics.}
Placement at standard message boundaries determines where a hotfix can operate, but not what it may do.
Matching only the type or fields of the current message would be simple, but insufficient for attacks that become evident only from an earlier message, a connection state, or an aggregate count. 
At the other extreme, allowing MNOs to execute arbitrary code would provide greater expressiveness but make hotfixes difficult to validate, audit, and execute safely.

\name adopts a \emph{stateful match-action abstraction} between these extremes. 
A condition identifies the protocol context that warrants intervention using the current message and, when needed, explicitly maintained RAN states. 
When the condition holds, one of three preventive actions is invoked, \textsc{Drop}, \textsc{Modify}, or \textsc{Release}, whose precise semantics are defined in \S\ref{sec:hostfix-construction}.
Conditions and actions are expressed using a closed set of declarative operations, rather than operator-supplied executable code, so that \name can validate their types, parameters, and resource use before enforcement.
These checks bound a rule's direct authority, while protocol-level review and testing remain necessary to establish the correctness of the particular mitigation.

This \emph{bounded} interface intentionally \emph{excludes more permissive operations}, such as arbitrary message synthesis, message scheduling, or new cryptographic processing.
Such operations can introduce protocol behavior beyond the processed message, require cooperation from endpoints outside the RAN, or make a rule's effects substantially harder to predict.
The selected interface thus favors predictable RAN-local intervention while retaining stateful context needed for coverage.

Section~\ref{sec:attack-corpus-analysis} empirically tests whether this bounded design retains useful coverage across the attack corpus.

\subsection{Architecture and Hotfix Lifecycle}
\label{sec:design:overview}

\begin{figure*}
    \centering
    \includegraphics[width=0.95\linewidth]{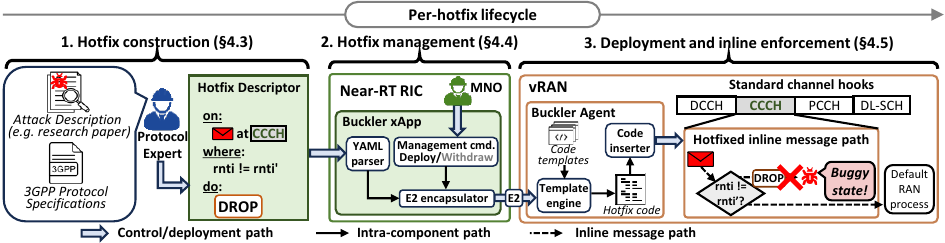}
    \vspace{-5pt}
    \caption{\name architecture and hotfix lifecycle.}
    \vspace{-10pt}
    \label{fig:design-overview}
\end{figure*}

Figure~\ref{fig:design-overview} presents the \name architecture and the
three phases in the lifecycle of each hotfix.  It also distinguishes the
control and deployment path used to construct, manage, and install a hotfix
from the inline message path on which the installed hotfix is enforced.

\para{1. Hotfix construction.}
A cellular-protocol expert analyzes an attack description together with the
relevant 3GPP specifications, expresses the mitigation as one or more bounded
stateful match-action rules, and serializes them in a declarative hotfix
descriptor.  Section~\ref{sec:hostfix-construction} defines this rule interface
and descriptor format.

\para{2. Hotfix management.}
The MNO submits the descriptor and subsequent deployment or withdrawal commands
to the \name xApp at the near-RT RIC.  The xApp parses the descriptor, manages
its lifecycle, and encapsulates the corresponding request for delivery over the
E2 interface.  Section~\ref{sec:hotfix-management} details these management
operations.

\para{3. Deployment and inline enforcement.}
The \name agent in the vRAN receives the descriptor, instantiates a handler from
predefined code templates, and loads it at the selected standard channel hook.
The handler then evaluates matching messages and applies the specified
\textsc{Drop}, \textsc{Modify}, or \textsc{Release} action directly on the
inline message path, before the default RAN processing can enter the vulnerable
state.  The MNO and the xApp therefore remain outside the per-message
enforcement path.  Section~\ref{sec:deployment-enforcement} describes this
process.

\subsection{Hotfix Construction}
\label{sec:hostfix-construction}

A hotfix developer is a cellular-protocol expert, either the MNO or any independent third-party, who assesses a vulnerability disclosure document and the relevant 3GPP specifications~\cite{3GPP-releases} to design an appropriate \name hotfix.
The disclosure may be a research paper, CVE report, vendor advisory, or another source that describes the attack procedure and its root cause in detail.
For the purpose of our corpus analysis in Section~\ref{sec:attack-corpus-analysis}, research papers are used for vulnerability disclosure documents as peer-reviewed papers are verified to detail the attack procedure, but the hotfix framework itself is independent of the disclosure format.

\para{Rule structure.}
A \name hotfix consists of one or more \emph{stateful match-action rules}.
For readability, we express each rule as
\begin{quote}
\underline{\textbf{on}} \emph{trigger} \quad
\underline{\textbf{where}} \emph{condition} \quad
\underline{\textbf{do}} \emph{action}.
\end{quote}
The \textbf{on} and \textbf{where} clauses together form the match side of the rule.
They define the attack detection condition that warrants intervention, such as adversary crafted messages or a vulnerable configurations.
The \textbf{do} clause specifies the preventive action to apply when the match succeeds.

\para{Trigger.}
The trigger names a standardized message type and the logical or transport channel on which it is handled, for example, an \texttt{RRCConnectionRequest} on CCCH.
This pair attaches the rule to a particular hook and determines when the rule is evaluated.
Evaluation occurs before the RAN's default processing of the matched message, allowing the action to prevent the unsafe transition rather than react afterward.

\para{Condition.}
The optional condition states when the triggering message requires intervention.
It may identify a malformed or adversary-crafted message, but it may also capture a vulnerable protocol context in which an otherwise well-formed message becomes unsafe.
For example, a condition can depend on prior messages, connection state, or a RAN-wide threshold.
A rule that must alter every instance of a message type can omit the condition.
When present, a condition may combine three forms of information:

\begin{packeditemize}
    \item[(i)] \textit{Current-message fields.}
    A rule may compare fields of the triggering message using exact, ordering, or membership predicates.

    \item[(ii)] \textit{Per-connection state.}
    A rule may test state derived from earlier messages, such as whether a connection has reached a particular procedure stage or whether a later message has cleared that state.

    \item[(iii)] \textit{RAN-wide aggregates.}
    A rule may test quantities maintained across connections, such as the number of connections currently stalled in a particular state.
\end{packeditemize}

The descriptor explicitly declares the data structures (e.g., maps, lists, or counters) needed to maintain the state used in these predicates and their corresponding state-updating rules.
State-updating rules record or remove entries in the data structures as relevant messages are observed but do not themselves intervene in protocol processing.

\para{Preventive action.}
When the condition holds, an intervention rule invokes one of the three actions established in \S\ref{sec:rationale}.
\textsc{Drop} withholds the matched message from further processing or transmission.
\textsc{Modify} rewrites selected fields of that message before processing or transmission continues.
\textsc{Release} gracefully terminates a connection with a UE and cleans up RAN-side  UE context; this may be the triggering connection or another connection selected by the condition.
State updates used to evaluate a condition are bookkeeping operations and are not additional preventive actions.

\para{Bounded descriptor.}
An operator serializes these rules in a high-level YAML configuration file we refer to as the hotfix descriptor.
The hotfix descriptor contains only declared state objects and a finite sequence of supported field, comparison, lookup, update, and intervention operations.
It admits neither loops nor operator-supplied function calls or executable code.
This closed vocabulary allows the xApp and the RAN-side runtime to reject malformed parameters and unsupported message fields or operations before installation.
The example descriptors of hotfixable attacks are available in our artifacts.

\subsection{Hotfix Management}
\label{sec:hotfix-management}
The \name xApp manages the lifecycle of a hotfix from the near-RT RIC.
It exposes a REST API through which an MNO issues hotfix management commands, and communicates with the \name agent at each vRAN over the standardized E2 interface~\cite{oran-e2ap}.
To deploy a hotfix, the MNO submits a YAML descriptor to the xApp endpoint.
The xApp parses the descriptor and populates a \texttt{RIC Control Request} with the declared rule components.
The request is ASN.1 encoded and delivered to the \name agent at the target vRAN, which responds with a \texttt{RIC Control Acknowledgment} carrying a one-byte hotfix identifier, or with a \texttt{RIC Control Failure} if it cannot instantiate the hotfix.
To withdraw a hotfix, the MNO issues a request specifying its identifier, which the xApp relays in a further \texttt{RIC Control Request}.
Upon receiving this message, the agent unloads the handler and releases the state the hotfix maintained.
This xApp-based hotfix management is performed entirely through O-RAN compliant interfaces, allowing an MNO to deploy and withdraw hotfixes uniformly across a multi-vendor RAN deployment.

\subsection{Deployment and Inline Enforcement}
\label{sec:deployment-enforcement}
Two components at the vRAN enforce hotfixes at runtime.
The \name agent constructs the hotfix code from a descriptor and installs it, and the standard channel hooks are the insertion points at which that code runs.

\para{\name agent.}
The agent holds a set of predefined code templates covering the message handling operations of the bounded vocabulary, namely message parsing, field extraction, state access, comparisons, state updates, and the three preventive actions.
The templating engine selects the templates a descriptor calls for and instantiates them with the declared parameters to produce hotfix code for the target channel and message type.
As hotfix code is assembled only from these templates, arbitrary code and operations cannot be installed into the RAN through a descriptor, which is what enforces the bounded hotfix semantics.
The code inserter then compiles this code and loads it at the hook the descriptor specifies.
The inserter adopts the eBPF-based approach of Janus~\cite{2023-MOBICOM-Janus} and TeleRAN~\cite{2026-FutureG-TeleRAN}, which demonstrated dynamic code insertion in vRAN software, allowing hotfixes to be installed and removed without manual code integration or process restarts.

\para{Standard channel hooks.}
Hooks are placed at the standardized channel boundaries, where the messages crossing each channel are exposed to the hotfix code loaded there.
The hotfix code intercepts a standard message arriving at the hook before the RAN's default processing of it begins, and applies the checks and modifications the hotfix defines.
Enforcing hotfix rules at these hooks rather than at the near-RT RIC eliminates the control-loop round trip and enables our inline preventive actions.
Table~\ref{table:hooks} enumerates the candidate hook locations, namely the standardized logical and transport channels~\cite{3GPP-TS38.321,3GPP-TS38.331}.
While the message types each hook exposes are standardized, the code-level changes needed to realize it vary with a given vRAN's structure and functional split, and several channels may share a single insertion site.
The resulting burden on vendors is nevertheless small, as Section~\ref{sec:attack-corpus-analysis} reports how few of these hooks the hotfixes we construct require and Section~\ref{sec:implementation} measures the lines of code needed to realize them.

\begin{table}[t]
\caption{Standard channel boundaries available for \name hooks.}
\vspace{-10pt}
\label{table:hooks}
\renewcommand{\arraystretch}{1.1}
\centering
\resizebox{\columnwidth}{!}{%
\begin{tabular}{@{} l l l l @{}}
\hline
\textbf{Channel Type} & \textbf{Channel} & \textbf{Representative Location} & \textbf{Scope} \\
\hline
\multirow{7}{*}{Logical} & BCCH & CU-C (RRC <-> MAC) & Broadcast messages \\
 & PCCH & CU-C (RRC <-> MAC) & Paging messages \\
 & CCCH & CU-C (RRC <-> RLC) & Initial connection setup \\
 & DCCH & CU-C (RRC <-> PDCP) & UE session control \\
 & MCCH & CU-C (RRC <-> MAC) & Multicast control \\
 & DTCH & CU-U (Core <-> PDCP) & User plane messages \\
 & MTCH & CU-U (Core <-> PDCP) & Multicast user plane messages \\
\hline
\multirow{5}{*}{Transport} & UL/DL-SCH & DU (MAC <-> PHY) & MAC control elements \\
 & RACH & DU (MAC <-> PHY) & Random access preamble \\
 & BCH & DU (MAC <-> PHY) & Broadcast messages \\
 & PCH & DU (MAC <-> PHY) & Paging messages \\
 & MCH & DU (MAC <-> PHY) & Multicast data \\
\hline
\end{tabular}
}
\vspace{-10pt}
\end{table}
\section{Corpus-Wide Design Validation}
\label{sec:attack-corpus-analysis}

Section~\ref{sec:design} deliberately bounds both where a \name hotfix may intervene and what it may do.  
Whether these restrictions still leave enough control to prevent attacks is ultimately an empirical question.  
We answer that question against a corpus of L2/L3 attacks published over the past decade.  
Specifically, we ask: (1) which attacks provide a preventative intervention point at the legitimate RAN, (2) which of those attacks admit a concrete hotfix using \name's bounded interface, (3) which conditions, actions, and channel hooks the resulting hotfixes require, and (4) which capabilities are missing in the unsupported cases.
We first construct the corpus and define our analysis procedure, then present three representative hotfix constructions before reporting the corpus-wide results.

\subsection{Corpus Construction}
\label{subsec:corpus}

We construct the corpus using a venue- and year-bounded search.  
We examine the proceedings from 2016 through 2025 of four top-tier security venues (IEEE S\&P, USENIX Security, ACM CCS, and NDSS) and two established venues for mobile systems (MobiCom and WiSec).  
We select papers whose titles contain ``4G'', ``LTE'', ``5G'', or ``cellular'', yielding an initial set of 86 papers.  
A detailed paper-by-paper breakdown and the complete analysis of hotfixable attacks appear in Appendix~\ref{appendix:vulnerability-analysis}.

We next apply paper-level filters.  
First, we retain 59 \emph{attack papers}, defined as papers that demonstrate at least one concrete attack against a deployed cellular network.  
We exclude those on defense frameworks, protocol enhancement proposals, and measurement studies that do not disclose a concrete new vulnerability.  
Second, we retain 37 papers whose attacks exploit behavior rooted in the 3GPP specification, excluding 20 papers on implementation flaws (e.g., memory corruption or protocol noncompliance) and two on operator misconfiguration (e.g., insecure algorithm selection).  
Finally, we exclude 14 papers whose vulnerabilities lie in core-internal protocols (e.g., GTP or SBI) or application-layer protocols (e.g., IMS or SIP), outside the L2/L3 message-processing scope of this work.  
The remaining 23 papers report 64 distinct L2/L3 protocol attacks.  
When one paper reports multiple attacks, we analyze them separately because they may expose different protocol preconditions and require different interventions.

\subsection{Analysis Methodology}
\label{subsec:corpus-method}
Corpus inclusion establishes that an attack targets standard L2/L3 protocol behavior, but does not establish that the operator's legitimate RAN has an opportunity to prevent it.  
Thus, we further examine each included attack along two dimensions: RAN-side intervenability and \name hotfixability.

\para{RAN-intervenable attacks.}
For each attack, we identify the earliest \emph{intervention event}: a standard L2/L3 message event that occurs before the attack takes effect and at which changing the RAN's behavior can eliminate a necessary attack precondition.  
The event may be an adversary-crafted message received by the RAN, a benign but vulnerable message transmitted by the RAN, or an otherwise ordinary message whose occurrence makes an aggregate RAN state unsafe.  
We classify an attack as \emph{RAN-intervenable} only when the RAN observes and handles this event and can act before the attack causes its intended effect.
This criterion is stronger than standard attack detection: inferring an attack occurrence seeing evidence after the effect, or observing side-effects related to the attack, does not provide a preventative intervention point.

The criterion is evaluated per attack, not solely from the adversary type in
Figure~\ref{fig:threat-model}.  
The four adversary types included in our threat model can all participate in attacks whose messages reach the legitimate RAN, but this does not mean that every attack they mount is RAN-intervenable. 
For example, an MitM or signal injector adversary may inject or overshadow a downlink message directly to a UE without giving the legitimate RAN an opportunity to intercept it.  
Such attacks fail the intervenability criterion even when the broader adversary type remains within our threat model.

\para{\name-hotfixable attacks.}
For every RAN-intervenable attack, we search for a concrete stateful match-action rule that uses only the operations in Section~\ref{sec:design}.
We call an attack \emph{\name-hotfixable} when the constructed rule satisfies three conditions.  
First, its trigger is available at one of \name's standard channel hooks and its condition uses only the triggering message and explicitly maintained RAN state.  
Second, applying \textsc{Drop}, \textsc{Modify}, or \textsc{Release} at that point removes a necessary precondition before the attack takes effect.  
Third, the rule preserves compliant protocol operation or confines any side effect to an explicit, bounded recovery cost appropriate for a temporary mitigation.

Our analysis records, for each attack, its necessary protocol precondition, the earliest intervention event, the required message fields and state, the preventative action, and the effect on compliant executions.  
If the original disclosure, such as the mitigation section of the academic paper that introduced the attack, proposes a countermeasure enforceable within \name, we adopt that countermeasure.  
Otherwise, we search for a RAN-local alternative. 
When we find none, despite our best effort, we record the capability required by the most direct countermeasure, such as endpoint-generated integrity protection or confidentiality.  
Appendix~\ref{appendix:vulnerability-analysis} provides the resulting argument for every attack, allowing the classifications in the main table to be checked against the attack description and the relevant 3GPP procedure.

This procedure establishes coverage \emph{constructively}.  
A successful case is witnessed by a concrete rule, whereas an unsuccessful search does not prove that no alternative exists. 
Accordingly, the number of hotfixes we report is a lower bound on the coverage achievable with \name's interface, not a claimed maximum.
Section~\ref{sec:discussion} discusses this interpretation and its implications for assisted hotfix construction.

\begin{table*}[t!]
\centering
\caption{Analysis of the 43 in-scope, RAN-intervenable L2/L3 attacks.}
\vspace{-5pt}
\label{table:one-big-beautiful-table}
\resizebox{\linewidth}{!}{%
\begin{tabular}{l l l c c c c c}
\hline
\textbf{No.} & \textbf{Vulnerability} & \textbf{Paper} & \textbf{Adversary} & \textbf{Req.\ Capability} & \textbf{Detectable} & \textbf{Hotfixable} & \textbf{Channel Hook} \\
\hline

1 & Selective service denial & \multirow{2}{*}{Shaik, et al.\ (NDSS'16)~\cite{2016-NDSS-Practical}} & MitM & \textsc{Modify} & $\bigcirc$ & $\bigcirc$ & DCCH \\
2 & TMSI inference w/ paging & & PE/UE & \textsc{Encrypt} & $\bigcirc$ & $\times$ & --- \\
\hline

3 & Auth.\ sync.\ failure attack & Hussain, et al.\ (NDSS'18)~\cite{2018-NDSS-LTEINSPEC} & UE & \textsc{Drop}$^{\dagger}$ & $\bigcirc$ & $\bigcirc$ & CCCH, DCCH \\
\hline

4 & Session confusion & Cremers, et al.\ (NDSS'19)~\cite{2019-NDSS-SESSIONCONF} & UE & \textsc{Drop}$^{\dagger}$ & $\bigcirc$ & $\bigcirc$ & DCCH \\
\hline

5 & Paging timing correlation (TORPEDO) & \multirow{1}{*}{Hussain, et al.\ (NDSS'19)~\cite{2019-NDSS-IMSICrack}} & PE/UE & \textsc{Modify} & $\bigcirc$ & $\bigcirc$ & PCCH \\
\hline

6 & BTS resource depletion & \multirow{4}{*}{Kim, et al.\ (S\&P'19)~\cite{2019-SP-LTEFUZZ}} & UE & \textsc{Release}$^{\dagger}$ & $\bigcirc$ & $\bigcirc$ & CCCH, DCCH \\
7 & Blind DoS (RRC spoofing) & & UE & \textsc{Drop}$^{\dagger}$ & $\bigcirc$ & $\bigcirc$ & CCCH, DCCH \\
8 & Remote de-registration & & UE & \textsc{Drop} & $\bigcirc$ & $\bigcirc$ & DCCH \\
9 & SMS phishing & & UE & \textsc{Drop} & $\bigcirc$ & {$\bigcirc$\rlap{$^{\ast}$}} & DCCH \\
\hline

10 & User data manipulation (aLTEr) & Rupprecht, et al.\ (S\&P'19)~\cite{2019-SP-ALTER} & MitM & \textsc{Sign} & $\times$ & $\times$ & --- \\
\hline

11 & Battery draining & Shaik, et al.\ (WiSec'19)~\cite{2019-WISEC-Exposecap} & MitM & \textsc{Sign} & $\times$ & $\times$ & --- \\
\hline

12 & Signaling storm (fake paging) & Yang, et al.\ (USENIX'19)~\cite{2019-USENIX-SigOver} & SI & \textsc{Sign} & $\triangle$ & $\times$ & --- \\
\hline

13 & NAS counter reset & \multirow{5}{*}{Hussain, et al.\ (CCS'19)~\cite{2019-CCS-5GReasoner}} & MitM & \textsc{Drop}$^{\dagger}$ & $\bigcirc$ & $\bigcirc$ & DCCH \\
14 & Exposing NAS sequence number & & PE & \textsc{Encrypt} & $\bigcirc$ & $\times$ & --- \\
15 & Neutralizing TMSI refreshment & & MitM & \textsc{Ack. Del.} & $\triangle$ & $\times$ & --- \\
16 & Installing null cipher/integrity & & MitM & \textsc{Release} & $\bigcirc$ & $\bigcirc$ & DCCH \\
17 & Exposing TMSI / paging occasion & & MitM & \textsc{Ack. Del.} & $\triangle$ & $\times$ & --- \\
\hline

18 & Uplink impersonation & \multirow{2}{*}{Rupprecht, et al.\ (NDSS'20)~\cite{2020-NDSS-IMP4GT}} & MitM & \textsc{Sign} & $\times$ & $\times$ & --- \\
19 & Downlink impersonation & & MitM & \textsc{Sign} & $\times$ & $\times$ & --- \\
\hline

20 & Paging storm & Fang, et al.\ (WiSec'20)~\cite{2020-WISEC-Botnet} & UE & \textsc{Sign} & $\triangle$ & $\times$ & --- \\
\hline

21 & Keystream reuse (bearer ID reuse) & \multirow{2}{*}{Rupprecht, et al.\ (USENIX'20)~\cite{2020-USENIX-ReVoLTE}} & UE & \textsc{Modify} & $\bigcirc$ & {$\bigcirc$\rlap{$^{\ast}$}} & DCCH \\
22 & Missing media plane encryption & & PE & \textsc{Encrypt} & $\times$ & $\times$ & --- \\
\hline

23 & Authentication abort w attach & \multirow{2}{*}{Chen, et al.\ (S\&P'21)~\cite{2021-SP-Bookworm}} & UE & \textsc{Drop}$^{\dagger}$ & $\bigcirc$ & $\bigcirc$ & DCCH \\
24 & Authentication abort w detach & & UE & \textsc{Drop}$^{\dagger}$ & $\bigcirc$ & $\bigcirc$ & DCCH \\
\hline

25 & SUCI-Catcher & Chlosta, et al.\ (WiSec'21)~\cite{2021-WISEC-SUCI} & MitM & \textsc{Drop} & $\bigcirc$ & $\bigcirc$ & DCCH \\
\hline

26 & CA side-channel location tracking & Lakshmanan, et al.\ (USENIX'21)~\cite{2021-SECURITY-CASideChannel} & PE & \textsc{Modify} & $\bigcirc$ & $\bigcirc$ & DL-SCH \\
\hline

27 & Radio resource draining & \multirow{6}{*}{Tan, et al.\ (MobiCom'21)~\cite{2021-Mobicom-IoTAttack}} & SI & \textsc{Drop}$^{\dagger}$ & $\bigcirc$ & $\bigcirc$ & UL-SCH \\
28 & Prolonged packet delivery & & SI & \textsc{Sign} & $\triangle$ & $\times$ & --- \\
29 & Flexible throughput limiting & & SI & \textsc{Sign} & $\triangle$ & $\times$ & --- \\
30 & Device localization & & SI & \textsc{Encrypt} & $\bigcirc$ & $\times$ & --- \\
31 & Packet delivery loop & & SI & \textsc{Sign} & $\triangle$ & $\times$ & --- \\
32 & Connection reset & & SI & \textsc{Sign} & $\triangle$ & $\times$ & --- \\
\hline

33 & DCI side channel & Bae, et al.\ (USENIX'22)~\cite{2022-USENIX-Watching} & PE & \textsc{Encrypt} & $\bigcirc$ & $\times$ & --- \\
\hline

34 & Passive UE localization & \multirow{2}{*}{Kotuliak, et al.\ (USENIX'22)~\cite{2022-USENIX-LTRACK}} & PE & \textsc{Encrypt} & $\bigcirc$ & $\times$ & --- \\
35 & Downlink IMSI extraction & & SI & \textsc{Sign} & $\times$ & $\times$ & --- \\
\hline

36 & Downlink DoS & \multirow{3}{*}{Erni, et al.\ (MobiCom'22)~\cite{2022-MobiCom-Adaptover}} & SI & \textsc{Sign} & $\triangle$ & $\times$ & --- \\
37 & Uplink DoS & & SI & \textsc{Modify}$^{\dagger}$ & $\bigcirc$ & $\bigcirc$ & DCCH \\
38 & Uplink IMSI extractor & & SI & \textsc{Modify}$^{\dagger}$ & $\bigcirc$ & {$\bigcirc$\rlap{$^{\ast}$}} & DCCH \\
\hline

39 & Duplicate emergency attach & Hu, et al.\ (MobiCom'22)~\cite{2022-Mobicom-Emergency} & UE & \textsc{Drop}$^{\dagger}$ & $\bigcirc$ & $\bigcirc$ & DCCH \\
\hline

40 & Deletion of allowed CAG list & \multirow{3}{*}{Al Ishtiaq, et al.\ (USENIX'24)~\cite{2024-USENIX-Hermes}} & MitM & \textsc{Drop}$^{\dagger}$ & $\bigcirc$ & $\bigcirc$ & DCCH \\
41 & Energy depletion w/ RRC Setup & & SI/MitM & \textsc{Sign} & $\triangle$ & $\times$ & --- \\
42 & NAS COUNT update attack & & MitM & \textsc{Sign} & $\times$ & $\times$ & --- \\
\hline

43 & Multi-stage downgrade & Luo, et al.\ (USENIX'25)~\cite{2025-USENIX-SNI5GECT} & SI & \textsc{Sign} & $\triangle$ & $\times$ & --- \\
\hline 
\footnotesize{\newline}
\end{tabular}
}
{\footnotesize
Detectable: $\bigcirc$ = Preventive detection (25) \quad $\triangle$ = Only postmortem detection (11) \quad $\times$ = Undetectable (7) \\
Hotfixable: $\bigcirc$ = Hotfixable from the RAN (20) \quad
$\times$ = Not hotfixable (23) \\
$^{\ast}$ Root cause addressed by a later specification change \\
$^{\dagger}$ Crafted rule: newly constructed hotfixes in this work \\
Adversary: MitM = man-in-the-middle relay, UE = malicious UE, SI = signal injector, PE = passive eavesdropper. \\
}
\end{table*}

\subsection{Representative Hotfix Constructions}
\label{subsec:hotfix-examples}

We illustrate the analysis with hotfixes to three attacks that collectively showcase all three preventative actions and different forms of protocol context.  
Blind DoS uses per-connection history and \textsc{Drop}, BTS resource depletion uses a RAN-wide aggregate and \textsc{Release}, and the carrier-aggregation side channel applies \textsc{Modify} unconditionally to a vulnerable downlink message.  
For brevity, we omit the complete attack procedures and their impacts and focus on how the hotfixes are constructed.  
The original attack papers cited in Table~\ref{table:one-big-beautiful-table} provide those details.

\para{Blind DoS.}
A spoofed \texttt{RRCConnectionRequest} carries a victim's TMSI under a new RNTI.  
If the RAN treats the request as the victim reconnecting, it preempts
the victim's active radio connection.  
The attack therefore depends on the RAN accepting a TMSI that is already bound to a different active RNTI.  
The hotfix maintains identifiers bound to active connections from prior connection setup and release messages and drops conflicting setup requests:

\begin{quote}
\underline{\textbf{on}} \texttt{RRCConnectionRequest(tmsi, rnti)}\\
\underline{\textbf{where}} \textbf{seen} \texttt{SetupComplete(tmsi, rnti$'$)}
  \textbf{and} \texttt{rnti$' \neq$ rnti}
  \textbf{and not seen} \texttt{Release(rnti$'$)}
  \textbf{since that} \texttt{SetupComplete}\\
\underline{\textbf{do}} \textbf{drop}
\end{quote}

Dropping the conflicting request prevents the RAN from preempting the active connection.  
A compliant UE normally releases its previous connection before establishing another one.
Although UEs that could not gracefully release its connection due to unexpected radio link failures can match this rule, it can recover its connection promptly through a reattach.
The hotfix therefore trades a short recovery delay in this exceptional case for preventing an attacker from repeatedly preempting an unrelated UE.

\para{BTS resource depletion.}
An adversary opens many dummy connections that stop before authentication, exhausting the RAN RRC resources.
As every individual request is well formed, no single message distinguishes the attack. 
Instead, the hotfix counts connections that have completed RRC setup but have not proceeded with an authentication response or release.  
Once this aggregate reaches an operator-selected threshold, the rule releases the oldest stalled connection:

\begin{quote}
\underline{\textbf{on}} \texttt{SetupComplete(rnti)}\\
\underline{\textbf{where}} \textbf{count} \{\texttt{r} :
  \textbf{seen} \texttt{SetupComplete(r)}
  \textbf{and not seen} \texttt{AuthResponse(r)}
  \textbf{since that} \texttt{SetupComplete}\} \texttt{$\geq$ N}\\
\underline{\textbf{do}} \textbf{release} \texttt{oldest(r)}
\end{quote}

This rule bounds the unauthenticated state that an attacker can accumulate.
An unusually slow legitimate connection may also be released when the bound is reached, particularly during an ongoing attack, but it can retry while the RAN remains available.  
The threshold therefore exposes an explicit operator-controlled tradeoff between attack resilience and tolerance for legitimate setup delay.

\para{Carrier-aggregation side channel.}
A passive observer tracks a UE through the secondary-cell activation pattern carried in a benign downlink message from the RAN.  
Following the countermeasure proposed in the original disclosure~\cite{2021-SECURITY-CASideChannel}, the hotfix rewrites each activation bitmap to a randomized superset of the cells actually required:

\begin{quote}
\underline{\textbf{on}} \texttt{SCellActivation(bitmap)}\\
\underline{\textbf{do}} \textbf{modify}
  \texttt{bitmap := random\_superset(bitmap)}
\end{quote}

Randomization removes the stable activation pattern used for tracking while retaining every cell required for service.  
Activating additional cells can increase UE energy consumption, so the operator can limit the degree or duration of randomization while the temporary hotfix is active.

\subsection{Corpus-Wide Results}
\label{subsec:corpus-results}

Table~\ref{table:one-big-beautiful-table} summarizes our attack-by-attack analysis of the 43 in-scope, RAN-intervenable attacks, including the required capability, detectability, hotfixability, and channel hook.
We use these to quantify the RAN-side intervention boundary and hotfix coverage, then examine bounded semantics, extensibility, and the capabilities missing from unsupported attacks.

\para{RAN-side intervention boundary.}
Of the 64 L2/L3 attacks, 43 provide a preventative intervention point at the
legitimate RAN.  
The remaining 21 fail this structural criterion.  
They include attacks conducted entirely through a standalone fake base station and attacks whose decisive adversarial transmission reaches the UE without
passing through the RAN.  
Their exclusion is therefore {\em not} a consequence of \name's chosen rule language or action set, but rather stems from the architectural limitations of a RAN-side defense.
The 43 RAN-intervenable attacks form the candidates against which we evaluate \name's bounded design.

\para{Hotfix coverage.}
We construct a \name hotfix for 20 of the 43 candidates, or 46.5\% of the RAN-intervenable set.
We distinguish these constructions by whether the mitigation proposed in the original disclosure can be translated directly into a \name rule.  
This direct translation is possible for eight attacks.  
For the other 12, the suggested mitigation cannot be mapped to a \name rule, so we identify a different RAN-local intervention.  
These newly constructed hotfixes are marked $\dagger$ in Table~\ref{table:one-big-beautiful-table}.  
The successful cases span all four adversary types in our threat model and include availability, integrity, and privacy attacks.  
Thus, coverage is not confined to one protocol procedure or one class of attacker.
{
For each of the 20 hotfixable attacks, we inspect the relevant specifications to determine whether a remediation has since been applied.
Three attacks, marked $\ast$ in Table~\ref{table:one-big-beautiful-table}, have received appropriate changes in the specifications, while the remaining 17 remains unfixed.
These 17 remain exploitable even against a fully spec-compliant network, further highlighting the value of \name.
Hotfixes for the three fixed attacks may also remain relevant in practice, as standardized security enhancements have been observed to remain absent from commercial deployments long after their release~\cite{2022-WiSec-CHIS, 2023-WiSec-EURS}.
}

\para{Bounded semantics and operational simplicity.}
All 20 hotfixes fit the same closed descriptor vocabulary.  
Twelve use \textsc{Drop}, six use \textsc{Modify}, and two use \textsc{Release}.
Their conditions draw only on the three context classes exposed in Section~\ref{sec:design}: current message fields, individual connection states, and aggregates maintained across connections.
None requires operator-supplied executable code, vendor-internal state, or an attack-specific extension to the action set.  
The corpus therefore tests more than whether three action names can be attached to known attacks: it shows that the complete rules, including the state needed to separate unsafe from compliant executions, remain within the bounded interface.

To quantify the structural complexity hidden behind this qualitative result, we additionally measure four properties of every descriptor: the number of message handlers, declared state objects, match predicates, and state-update operations. 
Across the 20 hotfixes, a descriptor contains a median of {\em 2} handlers and at most {\em 4}, a median of {\em 2} state objects and at most {\em 4}, and a median of {\em 2} predicates and {\em 2} state updates.  
These measurements make operational simplicity falsifiable: even within a closed language, descriptors that required dozens of handlers or deeply composed state predicates would remain difficult to inspect and operate.  
We report the completed distribution together with the released descriptors.

\begin{figure}[t]
    \centering
    \includegraphics[width=0.9\columnwidth]{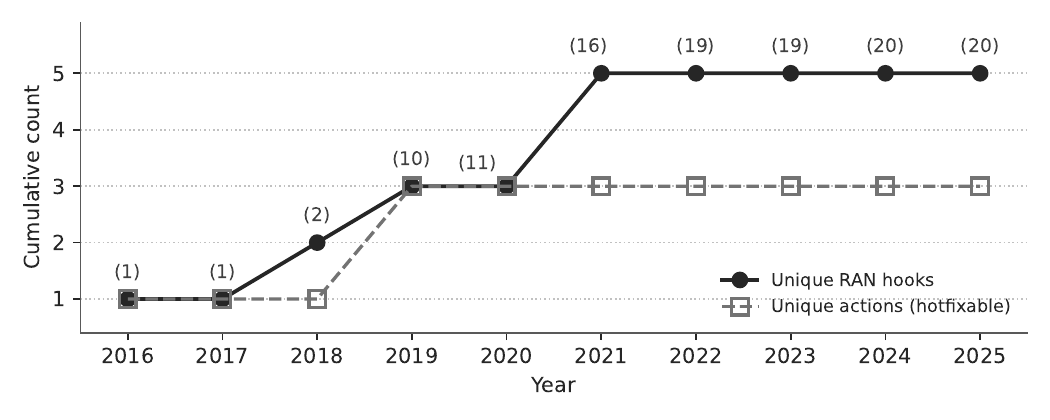}
    \vspace{-10pt}
    \caption{Cumulative number of distinct standard RAN channel hooks
    and actions required as papers enter the corpus in publication order.  The required
    hook set reaches five by 2021 and remains unchanged for the remainder of
    the corpus.}
    \vspace{-10pt}
    \label{fig:extensibility}
\end{figure}

\para{Extensibility across the corpus timeline.}
Extensibility requires newly disclosed attacks to reuse existing controls rather than induce another vendor modification.  
Figure~\ref{fig:extensibility} orders the papers chronologically and counts the distinct standard channel hooks required by the hotfixes available at each point.  
The first hotfix, in 2016, requires DCCH.  
CCCH is added in 2018, PCCH in 2019, and DL-SCH and UL-SCH in 2021.  
No subsequently published attack for which we construct a hotfix requires another hook.  
The hooks saturate at these five channels because they carry the messages the standard leaves unprotected, namely the initial-access and session-control signaling on CCCH and DCCH, and the paging and MAC control information on PCCH, DL-SCH, and UL-SCH.
Attacks that reach the RAN must exploit such messages, so newly disclosed ones continue to land on channels already instrumented.
The action set stabilizes even earlier: \textsc{Drop}, \textsc{Modify}, and \textsc{Release} have all appeared by 2019, and every later hotfix reuses them.
This retrospective saturation does not guarantee that no future attack will require a new hook or capability, but it provides evidence that the interface is reusable rather than a collection of per-attack controls.

\para{Limits exposed by unsupported attacks.}
For the remaining 23 RAN-intervenable attacks, the direct countermeasure we identify requires a capability intentionally excluded from \name. 
Fifteen require new origin authentication or integrity protection, six require confidentiality protection, and two requires verifiable delivery across an endpoint boundary.  
These operations cannot be supplied unilaterally by the RAN: a signature is ineffective unless the receiving UE or core function verifies it, encryption is ineffective unless the endpoint decrypts the field, and delivery verification requires corresponding endpoint behavior.
Allowing arbitrary RAN code would not remove these dependencies.  
The unsupported cases expose the architectural boundary of RAN-only hotfixing rather than calling for a more permissive rule language.

\para{Relation to 5G-Spector detection.}
To connect our results with the detection capability discussed in Section~\ref{sec:background}, we separately compare against the attacks detectable with 5G-Spector~\cite{2024-NDSS-5GSpector}.
As our corpus covers attacks beyond those analyzed in 5G-Spector, we determine for each of the 43 attacks whether its methodology of detection through RAN telemetry would apply, and record the verdict in the Detectable column of Table~\ref{table:one-big-beautiful-table}.
These verdicts follow from our interpretation of that methodology rather than from reported results.
Detectability is divided into three groups: 25 attacks detectable before the attack takes effect, 11 detectable only postmortem, and 7 undetectable from RAN telemetry.
Only the first group is of use to \name, since a preventive action must be taken before the attack takes effect.
Of these 25 attacks, 20 admit a \name hotfix.
This 80\% result has a different denominator from the corpus-wide 20-of-43 result: it asks how often an existing RAN-side detector could, in principle, be complemented with inline prevention.
The remaining five show that detection and hotfixability are not identical properties.
Each requires encrypting an exposed field, which the RAN cannot supply because the protection takes effect only once the receiver decrypts it.

The corpus provides evidence for the three properties in Section~\ref{sec:threat_model}.  
The bounded interface constructs hotfixes for a diverse subset of RAN-intervenable attacks, the required action and hook sets stabilize as the corpus grows, and every successful hotfix remains within a closed declarative vocabulary. 
\section{Implementation and Evaluation}
\label{sec:implementation}

We implement \name on two open-source RAN stacks and evaluate the following four questions left open by the design and corpus analysis. 
\begin{packeditemize}
    \item[\textbf{Q1.}] How many lines of source code changes are required to implement the five hooks and hotfix interface in each vRAN stack?
    \item[\textbf{Q2.}] Can the same operator-level hotfix abstraction be realized across the two stacks?
    \item[\textbf{Q3.}] Do representative hotfixes effectively prevent attacks end to end?
    \item[\textbf{Q4.}] How much runtime overhead does \name introduce?
\end{packeditemize}
We evaluate them using the representative blind DoS, BTS resource-depletion, and carrier-aggregation (CA) side-channel examples introduced in \S\ref{subsec:hotfix-examples}.

\subsection{Prototype and Testbed}
\label{subsec:prototype}

\para{Prototype components.}
Our prototype comprises an RIC containing the \name xApp, the \name agent, and a hook-instrumented vRAN. 
For the vRAN, we use srsRAN 23.11~\cite{srsRAN} and OpenAirInterface (OAI) v2.2.0~\cite{OAI}, and instrument them with eBPF hooks by utilizing Janus~\cite{2023-MOBICOM-Janus}.
The O-RAN Software Community near-RT RIC Release I~\cite{ORANSC} is used for the RIC, and a minimal \name xApp that exposes a REST endpoint for submitting YAML descriptors is implemented in Python. 
The xApp uses a modified RAN control E2SM~\cite{ORAN-RC} to deliver hotfix descriptor to the \name agent at the RAN over the standard E2 interface. 

The \name agent is implemented in Python using Jinja2~\cite{jinja2} to instantiate the hotfix code from predefined templates. 
It maps the bounded descriptor vocabulary to C handlers assembled from predefined blocks for parsing protocol fields, reading and updating declared state, and executing \textsc{Drop}, \textsc{Modify}, or \textsc{Release}. 
After receiving the descriptor, the handler code is compiled and loaded at the corresponding hook through Janus~\cite{2023-MOBICOM-Janus}.

\para{Hook-instrumented vRANs.}
We integrate Janus into both vRAN stacks and instrument their DCCH, CCCH, PCCH, DL-SCH, and UL-SCH processing paths.
Both stacks are instrumented and evaluated in their LTE configurations, as the example attacks were more reliably reproduced, and carrier aggregation, which the CA side channel relies on, is supported only in the LTE mode of srsRAN.
This does not affect the generality of the hooks, since the instrumented channels are defined identically in LTE and 5G.
Porting \name to a 5G deployment requires only updating the NR message definitions, while the hooks themselves remain unchanged.

\para{Testbed.}
Our end-to-end testbed runs the modified RAN, the O-RAN SC RIC, and an Open5GS core~\cite{OGS} on a 16-core Intel Core i9 workstation running Ubuntu 22.04. 
For the over-the-air attacks, a USRP B210 serves as the RAN's RF front end, a Samsung A20 smartphone is the victim UE, and a separate host equipped with a second B210 runs srsUE as the attacker UE. 
We reproduce the BTS resource-depletion and blind DoS attacks by modifying srsUE following their original attack procedure disclosure~\cite{2019-SP-LTEFUZZ}. 

\subsection{Integration Footprint and Cross-Implementation Portability}
\label{subsec:integration-portability}

Section~\ref{sec:attack-corpus-analysis} identifies the five channel hooks logically sufficient for the hotfixable attacks in our corpus. 
Here, we answer \textbf{Q1} by measuring the source-code footprint required to realize those hooks in the two vRAN implementations, and \textbf{Q2} by examining whether the same logical hooks and operator-facing abstraction map to both stacks.

\begin{table}[t]
    \centering
    \caption{Source-code footprint of the RAN-side substrate.}
    \vspace{-5pt}
    \label{table:implementation-footprint}
    \renewcommand{\arraystretch}{1.1}
    \begin{tabular}{lrrrr}
        \hline
        \textbf{RAN} & \textbf{Base} & \textbf{5 hooks} & \textbf{Full} \\
        \hline
        OAI     & 90  & 70 & 160 \\
        srsRAN  & 264 & 62 & 326  \\
        \hline
    \end{tabular}
    \vspace{-10pt}
\end{table}

\para{RAN integration footprint.}
Table~\ref{table:implementation-footprint} reports the lines of code (LoC) added to each RAN. 
The base Janus integration requires 90 LoC in OAI and 264 LoC in srsRAN. 
Declaring a channel hook requires 14 and 12 additional LoC on average, respectively. 
Implementing all five hooks on the two stacks requires 160 LoC in OAI and 326 LoC in srsRAN. 
The larger srsRAN footprint is primarily due to ASN.1 parsing helpers already available in OAI. 

\para{Cross-implementation portability.}
Although the two implementations differ syntactically and invoke different internal helpers, they dispatch the same standardized message classes and expose the raw payload and connection identifier at the same protocol boundary.
Consequently, the same five logical hooks and operator-facing hotfix abstraction map to both stacks without relying on common internal functions or data structures.
This structural correspondence answers \textbf{Q2}: the abstraction is portable across the two implementations, while each stack retains its own hook adapter.
{Appendix~\ref{appendix:portability-evaluation} extends this result to enforcement, loading the BTS resource depletion hotfix on OAI using the same descriptor as the srsRAN evaluation.}

\subsection{End-to-End Hotfix Effectiveness}
\label{subsec:end-to-end-effectiveness}

\begin{figure*}[t]
  \centering
  \begin{subfigure}[b]{0.32\textwidth}
    \centering
    \includegraphics[width=\linewidth]{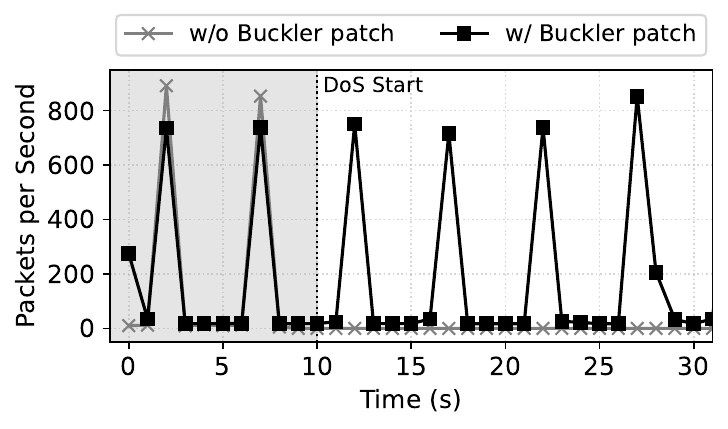}
    \caption{Blind DoS (\textsc{Drop})}
    \label{fig:blind-dos-traffic}
  \end{subfigure}
  \hfill
  \begin{subfigure}[b]{0.32\textwidth}
    \centering
    \includegraphics[width=\linewidth]{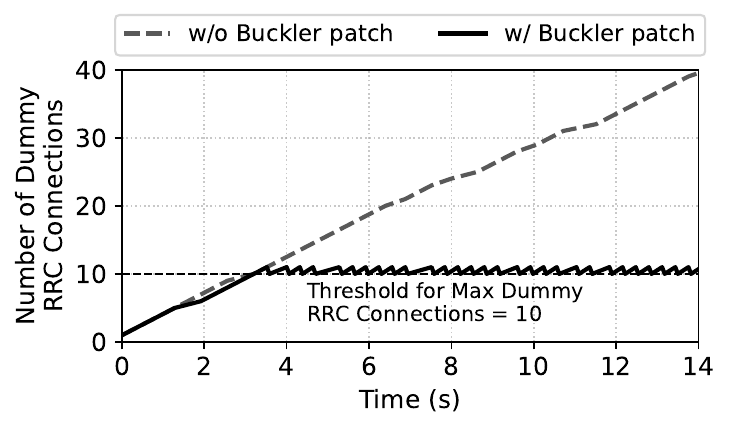}
    \caption{BTS resource depletion (\textsc{Release})}
    \label{fig:bts-depletion-evaluation}
  \end{subfigure}
  \hfill
  \begin{subfigure}[b]{0.32\textwidth}
    \centering
    \includegraphics[width=\linewidth]{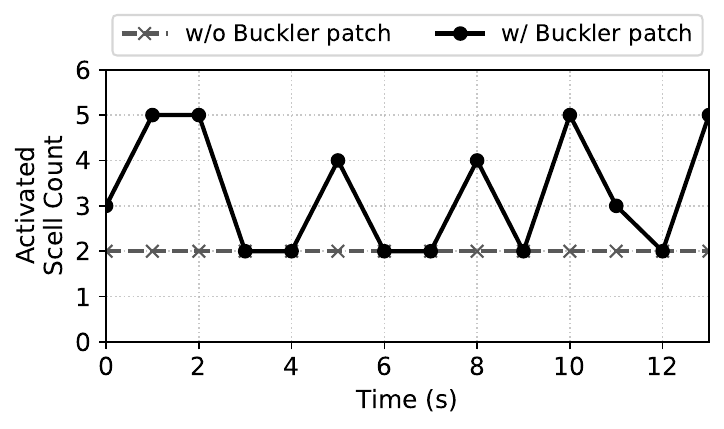}
    \caption{CA side channel (\textsc{Modify})}
    \label{fig:ca-side-channel-enforcement}
  \end{subfigure}
  \vspace{-5pt}
  \caption{Evaluation of hotfixes for three attacks: (a) blind DoS, (b) BTS resource depletion, and (c) the CA side channel. The three hotfixes demonstrate the preventive actions \textsc{Drop}, \textsc{Release}, and \textsc{Modify}, respectively. Every panel compares attack execution on two otherwise identical RANs, one without the hotfix (gray) and one with it loaded (black).}
  \vspace{-10pt}
  \label{fig:eval}
\end{figure*}

We next answer \textbf{Q3} by evaluating the representative hotfix for each attack. 
These cases demonstrate each of the preventive action available in \name, \textsc{Drop} is used in blind DoS, \textsc{Release} in BTS resource depletion, and \textsc{Modify} in the CA side channel attack.
The full hotfix descriptors for each attack are available in our artifacts.

\para{Blind DoS: suppressing a conflicting request.}
Our 117-LoC descriptor records active TMSIs and invokes \textsc{Drop} when a new \texttt{RRC Connection Request} presents a TMSI that is already active. 
We obtain the victim's TMSI from device debug output, configure the attacker's srsUE to connect using it, and stream continuous video to the victim so that any service interruption appears in its downlink trace.
Figure~\ref{fig:blind-dos-traffic} shows a representative run against otherwise identical unpatched and hotfixed RANs, with the attack starting at $t=10$ seconds. 

\para{BTS resource depletion: bounding attacker-created state.}
We modify the attacker's srsUE to reset its PHY parameters after receiving a \texttt{NAS Authentication Request}, to repeatedly create new RRC connections. 
Under the default Open5GS configuration, each dummy connection only times out after 15 s.
The 136-LoC descriptor tracks unauthenticated connections in first-in-first-out order and invokes \textsc{Release} on the oldest once the active connection count exceeds ten.
Figure~\ref{fig:bts-depletion-evaluation} shows dummy connections accumulating throughout the attack window without the hotfix and held below the threshold with it, demonstrating that our hotfix can clean up the dummy connections effectively.

\para{CA side channel: rewriting a vulnerable transmission.}
Following the original paper's mitigation~\cite{2021-SECURITY-CASideChannel}, the hotfix invokes \textsc{Modify} and replaces every secondary-cell activation bitmap to a randomized superset. 
This hotfix is tested under srsRAN's ZeroMQ RF emulation, the only mode that supports carrier aggregation.
Figure~\ref{fig:ca-side-channel-enforcement} shows the emitted bitmap activating a different number of secondary cells than the original command, validating interception and rewriting at the downlink transport-channel hook.

\subsection{Runtime and Operational Costs}
\label{subsec:runtime-operational-costs}

\begin{figure}[t]
    \centering
    \includegraphics[width=0.84\linewidth]{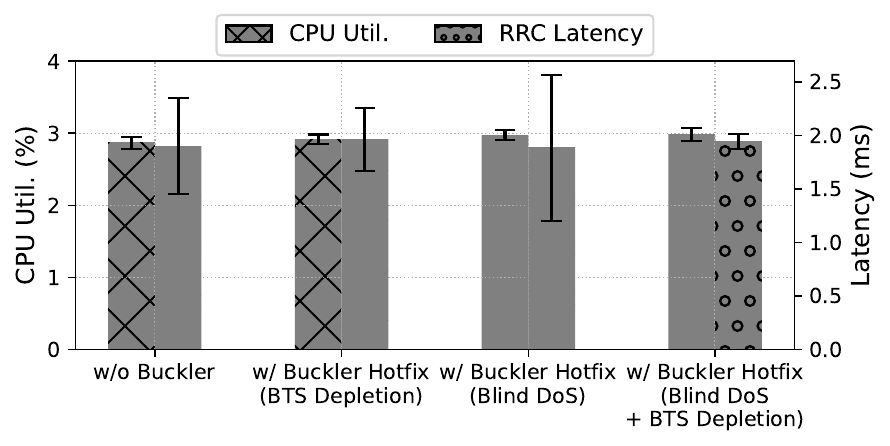}
    \vspace{-7pt}
    \caption{srsRAN CPU utilization and RRC response latency with the Blind DoS and BTS depletion hotfixes installed.}
    \vspace{-10pt}
    \label{fig:perf-overhead}
\end{figure}

We finally answer \textbf{Q4} by evaluating runtime and operational costs on the srsRAN testbed.
Unless stated otherwise, the measurements compare the same RAN build without a loaded hotfix against builds with the blind DoS or BTS resource-depletion hotfix active.

\para{Per-message CPU and response latency.}
Both hotfixes perform frequent state accesses while processing \texttt{RRC Connection Request} messages.
To stress these paths, we repeatedly attach and detach an srsUE and measure mean RAN-process CPU utilization over ten attach/detach cycles, repeat the measurement ten times for each configuration, and measure response latency for \texttt{RRC Connection Request} messages over 30 sequential attaches.

Figure~\ref{fig:perf-overhead} reports the results.
Relative to the unpatched baseline, CPU utilization increases by 3.55\% with the blind DoS hotfix and 1.57\% with the BTS resource-depletion hotfix, and mean RRC response latency increases by at most 3.37\%.
Even with both hotfixes loaded at once, CPU utilization increases by 3.80\%.
A stateful handler therefore adds modest processing cost even on the attach path.

\para{Installation and withdrawal.}
Over 30 load/unload cycles repeated ten times, installing a hotfix takes 356.9\,ms on average and withdrawing one takes 55.5\,ms.
Both are dominated by the underlying code insertion mechanism rather than by \name itself, and neither interrupts the messages passing through the affected hook.

Taken together, these measurements show that \name's inline enforcement imposes modest and sub-additive runtime cost, and that hotfixes can be installed and withdrawn on a live RAN within a fraction of a second.
\section{Discussion}
\label{sec:discussion}

\para{Constructive lower bound on hotfix coverage.}
\label{sec:constructive-coverage}
Our corpus analysis establishes hotfix coverage by construction.
For each of the 20 successful cases, we construct a concrete \name rule showing that the attack can be prevented within the bounded interface.
The converse does not follow: failing to find a rule for another attack does not prove that no such rule exists, because hotfix construction requires identifying both a RAN-controlled protocol precondition and a condition that separates unsafe execution from compliant behavior.
The 20 hotfixes should therefore be interpreted as a constructive lower bound on \name's coverage, not as its maximum.

The unsuccessful cases nevertheless clarify the present boundary.
For each, we report the most direct countermeasure we identified and the capability it requires.
Their concentration in endpoint-dependent authentication, confidentiality, and delivery verification explains why they fall outside a RAN-only design.
A different attack-specific insight could still reveal another RAN-local precondition and yield a hotfix without changing the interface.

\para{AI-assisted hotfix construction.}
\name does not automate the derivation of a hotfix from a vulnerability disclosure.
A protocol expert must examine the attack and relevant 3GPP procedures, identify an intervention point, and determine how the rule affects compliant executions.
Reducing this manual effort is a promising use of large language model (LLM) assistance.
An LLM could propose candidate triggers, conditions, actions, and state definitions in \name's descriptor language, but each candidate would still require protocol review and testing against compliant behavior.
The bounded vocabulary limits what an assisted construction can request, even though it does not by itself establish that the resulting rule is semantically correct.

\para{Limitations.}
First, our venue-, year-, and title-bounded corpus is systematic but not exhaustive, and the constructive methodology yields a lower bound rather than complete coverage of all L2/L3 attacks.
Second, we implement \name on two open-source RAN stacks in their LTE configurations.
The hooks follow standardized channel boundaries shared by LTE and 5G, but portability to proprietary stacks and end-to-end operation on a native 5G deployment remain to be demonstrated.
Third, the representative experiments validate enforcement of all three actions, but the current CA side-channel experiment demonstrates message rewriting rather than the resulting reduction in an observer's inference accuracy.
Finally, the bounded descriptor language constrains operator authority and makes rules more inspectable, but it cannot guarantee that a valid rule preserves all compliant executions; protocol review and testing by the hotfix developer and deploying MNO remain necessary before deployment.

\section{Related Work}
\label{sec:related-work}

\para{Security of the O-RAN platform.}
O-RAN introduces new security concerns through its open interfaces, RAN Intelligent Controllers (RICs), and third-party xApps.
The O-RAN Alliance Work Group 11 has documented these attack surfaces~\cite{ORAN-SEC,ORAN-RICSEC}, which have also been examined in recent academic work~\cite{2024-IEEE-xAppSec}.
Proposed defenses include fine-grained xApp authorization~\cite{2023-INFOCOM-xAppAuth} and zero-trust designs such as CMF~\cite{2023-IEEE-CONF} and ZTRAN~\cite{2024-IEEE-ZTRAN}.
Other studies test O-RAN's A1~\cite{2024-WiSec-OAITT}, E2~\cite{2024-USENIX-ORANalyst}, and open-fronthaul interfaces~\cite{2025-USENIX-fronthaul}.
Our threat model assumes that this platform is not compromised.
These efforts therefore protect the substrate on which \name depends, whereas \name uses trusted O-RAN programmability to mitigate known L2/L3 protocol vulnerabilities.
Its bounded descriptor interface limits the authority exposed to an operator, but does not replace authentication, authorization, or interface security for the O-RAN platform itself.

\para{Security through O-RAN programmability.}
Several systems use O-RAN programmability to improve visibility into cellular attacks.
5G-Spector~\cite{2024-NDSS-5GSpector} detects previously disclosed L3 attacks from security-focused RAN telemetry, Spotlight~\cite{2024-MOBICOM-Spotlight} applies machine learning to throughput KPIs for anomaly detection, and Dai et al.~\cite{2024-IEEE-ODET} demonstrate attack-detection pipelines using radio-signal data.
These systems show that operator applications can obtain security-relevant RAN observations, but do not provide general inline intervention on the corresponding protocol messages.
Janus~\cite{2023-MOBICOM-Janus} instead enables dynamic instrumentation of running RAN software and provides the runtime code-insertion substrate adopted by our prototype.
\name builds on these opportunities by exposing a fixed set of standard-message hooks and a bounded, stateful match-action interface for preventing known attacks before the vulnerable RAN behavior takes effect.
Our corpus analysis further tests whether that common interface, rather than an attack-specific detector or control, remains useful across a broad set of disclosed attacks.
\section{Conclusion}
\label{sec:conclusion}

In this work, we formulate operator-empowered hotfixing as an interim defense against known vulnerabilities in standard L2/L3 cellular protocol behavior.
Our framework, \name, enables an MNO to deploy temporary, local, and reversible hotfixes through reusable hooks at standardized RAN channel boundaries and to express them using a closed, stateful match-action interface.
Across a corpus of 64 protocol attacks, 43 provide a preventive intervention point at the legitimate RAN, and we construct \name hotfixes for 20 using the same three actions and five channel hooks.
We implement these hooks on srsRAN and OpenAirInterface with small, structurally similar changes, demonstrate end-to-end \textsc{Drop} and \textsc{Release} enforcement against representative availability attacks, and functionally validate \textsc{Modify} for a privacy attack.
Together, these results show that operator-empowered hotfixing can provide practical, portable attack prevention during the disclosure-to-deployment window, while the unsupported cases delineate the endpoint dependencies that a RAN-only defense cannot overcome.
\name complements, rather than replaces, standards- and vendor-led permanent remediation.

{\footnotesize
\bibliographystyle{plain}
\bibliography{main}

@misc{3GPP-TS38.331,
    title="{TS 38.331 v17.4.0: 	NR; Radio Resource Control (RRC); Protocol specification}",
    author="3GPP",
    year=2023
}

@misc{3GPP-TS38.321,
    title="{TS 38.321 v17.1.0: 5G; NR; Medium Access Control (MAC) protocol specification}",
    author="3GPP",
    year=2022
}

@inproceedings{2023-MOBICOM-Janus,
  title={{Taking 5G RAN Analytics and Control to a New Level}},
  author={Foukas, Xenofon and Radunovic, Bozidar and Balkwill, Matthew and Lai, Zhihua},
  booktitle={Proc. ACM MobiCom},
  year={2023}
}

@inproceedings{2024-NDSS-5GSpector,
  title={{5G-spector: An O-RAN Compliant Layer-3 Cellular Attack Detection Service}},
  author={Wen, Haohuang and Porras, Phillip and Yegneswaran, Vinod and Gehani, Ashish and Lin, Zhiqiang},
  booktitle={Proc. NDSS},
  year={2024}
}

@misc{oran-arch-site,
    author={{O-RAN Project}},
    title={{O-RAN Architecture Overview}},
    year=2022,
    howpublished={\url{https://docs.o-ran-sc.org/en/latest/architecture/architecture.html}},
}

@misc{3GPP-releases,
  author       = {{3GPP}},
  title        = {{3GPP Releases}},
  year         = {2025},
  howpublished = {\url{https://portal.3gpp.org/\#/55934-releases}},
}

@misc{Ettus,
  title        = {{USRP Hardware Driver and Manual}},
  author       = {{Ettus Research}},
  year         = {2025},
  howpublished = {\url{https://files.ettus.com/manual/index.html}},
}

@misc{BladeRF,
  title        = {{BladeRF Wiki}},
  author       = {{Nuand}},
  year         = {2023},
  howpublished = {\url{https://github.com/Nuand/bladeRF/wiki}},
}

@article{jinja2,
  title={Jinja2 documentation},
  author={Ronacher, Armin},
  journal={Welcome to Jinja2—Jinja2 Documentation (2.8-dev)},
  year={2008}
}

@misc{srsRAN,
  author       = {{Software Radio Systems}},
  title        = {{srsRAN\_4G}},
  year         = {2024},
  howpublished = {\url{https://github.com/srsran/srsRAN_4G}},
}

@misc{OAI,
  author       = {{OpenAirInterface Software Alliance}},
  title        = {{OpenAirInterface5G: Open Source 5G Software}},
  year         = {2024},
  howpublished = {\url{https://gitlab.eurecom.fr/oai/openairinterface5g}},
}

@misc{OGS,
  author       = {{Open5GS}},
  title        = {{Open5GS}},
  year         = {2024},
  howpublished = {\url{https://open5gs.org/}},
}

@article{2023-IEEE-CONF,
  title={{Conflict Mitigation Framework and Conflict Detection in O-RAN near-RT RIC}},
  author={Adamczyk, Cezary and Kliks, Adrian},
  journal={IEEE Communications Magazine},
  volume={61},
  number={12},
  pages={199--205},
  year={2023},
}

@article{2024-IEEE-ZTRAN,
  title={{ZTRAN: Prototyping Zero Trust Security xApps for Open Radio Access Network Deployments}},
  author={Abdalla, Aly S and Moore, Joshua and Adhikari, Nisha and Marojevic, Vuk},
  journal={IEEE Wireless Communications},
  volume={31},
  number={2},
  pages={66--73},
  year={2024},
}

@article{2024-IEEE-ODET,
  title={{Detection of Overshadowing Attack in 4G and 5G Networks}},
  author={Dai, Jiongyu and Saeed, Usama and Wang, Ying and Pan, Yanjun and Wang, Haining and Kornegay, Kevin T and Liu, Lingjia},
  journal={IEEE/ACM Transactions on Networking},
  year={2024},
  volume={32},
  number={6},
  pages={4615-4628},
}

@inproceedings{2024-USENIX-ORANalyst,
  title={{ORANalyst: Systematic Testing Framework for Open RAN Implementations}},
  author={Yang, Tianchang and Rashid, Syed Md Mukit and Ranjbar, Ali and Tan, Gang and Hussain, Syed Rafiul},
  booktitle={Proc. USENIX Security},
  pages={1921--1938},
  year={2024}
}

@inproceedings{2024-WiSec-OAITT,
  title={Security testing the o-ran near-real time ric \& a1 interface},
  author={Thimmaraju, Kashyap and Shaik, Altaf and Fl{\"u}ck, Sunniva and Mora, Pere Joan Fullana and Werling, Christian and Seifert, Jean-Pierre},
  booktitle={Proc. ACM WiSec},
  pages={277--287},
  year={2024}
}

@inproceedings{2025-USENIX-fronthaul,
  title={On the criticality of integrity protection in 5G fronthaul networks},
  author={Xing, Jiarong and Yoo, Sophia and Foukas, Xenofon and Kim, Daehyeok and Reiter, Michael K},
  booktitle={Proc. USENIX Security},
  pages={4463--4479},
  year={2024}
}

@inproceedings{2024-MOBICOM-Spotlight,
  title={{SpotLight: Accurate, explainable and efficient anomaly detection for Open RAN}},
  author={Sun, Chuanhao and Pawar, Ujjwal and Khoja, Molham and Foukas, Xenofon and Marina, Mahesh K and Radunovic, Bozidar},
  booktitle={Proc. ACM MobiCom},
  year={2024}
}

@inproceedings{2023-INFOCOM-xAppAuth,
  title={Securing 5g openran with a scalable authorization framework for xapps},
  author={Atalay, Tolga O and Maitra, Sudip and Stojadinovic, Dragoslav and Stavrou, Angelos and Wang, Haining},
  booktitle={IEEE INFOCOM 2023-IEEE Conference on Computer Communications},
  pages={1--10},
  year={2023},
  organization={IEEE}
}

@article{2024-IEEE-xAppSec,
  title={Security threats to xApps access control and E2 interface in O-RAN},
  author={Hung, Cheng-Feng and Chen, You-Run and Tseng, Chi-Heng and Cheng, Shin-Ming},
  journal={IEEE Open Journal of the Communications Society},
  volume={5},
  pages={1197--1203},
  year={2024},
  publisher={IEEE}
}

@misc{vran-samsung,
    title={{vRAN Value Proposition and Cost Modeling}},
    author={Young, Lee and Hyunjeong, Lee and Jai-Jin, Lim},
    month={Sep},
    year={2020},
    howpublished={Whitepaper: \url{https://images.samsung.com/is/content/samsung/assets/global/business/networks/insights/white-papers/vran-value-proposition-and-cost-modeling/vRAN-Value-Proposition-and-Cost-Modeling-Whitepaper.pdf}},
}

@misc{vran-nec,
    title={{Building an Open vRAN Ecosystem}},
    author={{NEC Corporation}},
    year={2020},
    howpublished={Whitepaper: \url{https://www.nec.com/en/global/solutions/5g/download/pdf/Building_an_Open_vRAN_Ecosystem_White_Papaer.pdf}},
}

@misc{vran-mavenir,
    title={{World’s First 5G SA Network Using Open vRAN on a Public Cloud}},
    author={Mavenir},
    year={2023},
    howpublished={Whitepaper: \url{https://www.mavenir.com/case-studies/mavenir-and-dish/}},
}

@misc{vran-rakuten,
    title={{Rakuten Claims Huge Edge Cloud, as Other Operators Follow Suit}},
    author={Caroline Gabriel},
    year={2019},
    howpublished={\url{https://rethinkresearch.biz/articles/
rakuten-claims-huge-edge-cloud-as-other-operators-follow-suit/}},
}

@inproceedings{2023-WiSec-EURS,
    author = {Lasierra, Oscar and Garcia-Aviles, Gines and Municio, Esteban and Skarmeta, Antonio and Costa-P\'{e}rez, Xavier},
    title = {{European 5G Security in the Wild: Reality versus Expectations}},
    year = {2023},
    booktitle = {Proc. ACM WiSec},
}

@inproceedings{2018-NDSS-LTEINSPEC,
  title={{LTEInspector: A Systematic Approach for Adversarial Testing of 4G LTE}},
  author={Hussain, Syed and Chowdhury, Omar and Mehnaz, Shagufta and Bertino, Elisa},
  booktitle={Proc. NDSS},
  year={2018}
}

@inproceedings{2019-CCS-5GReasoner,
    title={{5GReasoner: A Property-directed Security and Privacy Analysis Framework for 5G Cellular Network Protocol}},
    author={Hussain, Syed Rafiul and Echeverria, Mitziu and Karim, Imtiaz and Chowdhury, Omar and Bertino, Elisa},
    booktitle={Proc. ACM CCS},
    year={2019}
}

@inproceedings{2022-WiSec-CHIS,
    author = {Nie, Shiyue and Zhang, Yiming and Wan, Tao and Duan, Haixin and Li, Song},
    title = {{Measuring the Deployment of 5G Security Enhancement}},
    year = {2022},
    booktitle = {Proc. ACM WiSec},
}

@inproceedings{2019-SP-LTEFUZZ,
    title={{Touching the Untouchables: Dynamic Security Analysis of the LTE Control Plane}},
    author={Kim, Hongil and Lee, Jiho and Lee, Eunkyu and Kim, Yongdae},
    booktitle={Proc. IEEE S\&P},
    year={2019},
}

@inproceedings{2019-SP-ALTER,
    title={{Breaking LTE on Layer Two}},
    author={Rupprecht, David and Kohls, Katharina and Holz, Thorsten and P{\"o}pper, Christina},
    booktitle={Proc. IEEE S\&P},
    year={2019},
}

@inproceedings{2016-NDSS-Practical,
    title={{Practical Attacks Against Privacy and Availability in 4G/LTE Mobile Communication Systems}},
    author={Shaik, Altaf and Borgaonkar, Ravishankar and Asokan, N and Niemi, Valtteri and Seifert, Jean-Pierre},
    booktitle = {Proc. NDSS},
    year = {2016}
}

@inproceedings{2022-USENIX-Watching,
  title={{Watching the Watchers: Practical Video Identification Attack in LTE Networks}},
  author={Bae, Sangwook and Son, Mincheol and Kim, Dongkwan and Park, CheolJun and Lee, Jiho and Son, Sooel and Kim, Yongdae},
  booktitle={Proc. USENIX Security},
  year={2022}
}

@inproceedings{2019-NDSS-IMSICrack,
  title={{Privacy Attacks to the 4G and 5G Cellular Paging Protocols using Side Channel Information}},
  author={Hussain, Syed Rafiul and Echeverria, Mitziu and Chowdhury, Omar and Li, Ninghui and Bertino, Elisa},
  booktitle={Proc. NDSS},
  year={2019}
}

@inproceedings{2019-USENIX-SigOver,
  title={{Hiding in plain signal: Physical signal overshadowing attack on LTE}},
  author={Yang, Hojoon and Bae, Sangwook and Son, Mincheol and Kim, Hongil and Kim, Song Min and Kim, Yongdae},
  booktitle={Proc. USENIX Security},
  pages={55--72},
  year={2019}
}

@inproceedings{2022-MobiCom-Adaptover,
  title={{AdaptOver: Adaptive Overshadowing Attacks in Cellular Networks}},
  author={Erni, Simon and Kotuliak, Martin and Leu, Patrick and Roeschlin, Marc and Capkun, Srdjan},
  booktitle={Proc. ACM MobiCom},
  year={2022}
}

@inproceedings{2024-USENIX-Hermes,
  title={{Hermes: Unlocking Security Analysis of Cellular Network Protocols by Synthesizing Finite State Machines from Natural Language Specifications}},
  author={Ishtiaq, Abdullah\_Al and Das, Sarkar\_Snigdha S and Rashid, Syed\_Md M and Ranjbar, Ali and Tu, Kai and Wu, Tianwei and Song, Zhezheng and Wang, Weixuan and Akon, Mujtahid and Zhang, Rui and others},
  booktitle={Proc. USENIX Security},
  year={2024}
}

@inproceedings{2025-USENIX-SNI5GECT,
  title={$\{$SNI5GECT$\}$: A Practical Approach to Inject $\{$aNRchy$\}$ into 5G $\{$NR$\}$},
  author={Luo, Shijie and Garbelini, Matheus and Chattopadhyay, Sudipta and Zhou, Jianying},
  booktitle={Proc. USENIX Security},
  pages={5385--5404},
  year={2025}
}

@inproceedings{2021-SP-Bookworm,
  title={Bookworm game: Automatic discovery of LTE vulnerabilities through documentation analysis},
  author={Chen, Yi and Yao, Yepeng and Wang, XiaoFeng and Xu, Dandan and Yue, Chang and Liu, Xiaozhong and Chen, Kai and Tang, Haixu and Liu, Baoxu},
  booktitle={Proc. IEEE S\&P},
  pages={1197--1214},
  year={2021},
  organization={IEEE}
}

@inproceedings{2019-WISEC-Exposecap,
  title={New vulnerabilities in 4G and 5G cellular access network protocols: exposing device capabilities},
  author={Shaik, Altaf and Borgaonkar, Ravishankar and Park, Shinjo and Seifert, Jean-Pierre},
  booktitle={Proc. ACM WiSec},
  pages={221--231},
  year={2019}
}

@inproceedings{2022-Mobicom-Emergency,
  title={Uncovering insecure designs of cellular emergency services (911)},
  author={Hu, Yiwen and Chen, Min-Yue and Tu, Guan-Hua and Li, Chi-Yu and Wang, Sihan and Shi, Jingwen and Xie, Tian and Xiao, Li and Peng, Chunyi and Tan, Zhaowei and others},
  booktitle={Proc. ACM MobiCom},
  pages={703--715},
  year={2022}
}

@inproceedings{2021-Mobicom-IoTAttack,
  title={Data-plane signaling in cellular IoT: Attacks and defense},
  author={Tan, Zhaowei and Ding, Boyan and Zhao, Jinghao and Guo, Yunqi and Lu, Songwu},
  booktitle={Proc. ACM MobiCom},
  pages={465--477},
  year={2021}
}

@inproceedings{2021-WISEC-SUCI,
  title={{5G SUCI-Catchers: Still catching them all?}},
  author={Chlosta, Merlin and Rupprecht, David and P{\"o}pper, Christina and Holz, Thorsten},
  booktitle={Proc. ACM WiSec},
  year={2021}
}

@misc{cvd-slic,
  title        = {{CVD-2020-0040}: Stealthy Location Identification Attack exploiting Carrier Aggregation},
  author       = {{GSMA}},
  year         = {2020},
  howpublished = {\url{https://www.gsma.com/solutions-and-impact/technologies/security/gsma-mobile-security-research-acknowledgements/}},
}

@misc{ORANSC,
    author={{Open RAN Alliance Software Community}},
    title={O-RAN SC I Release},
    year={2023},
    howpublished={\url{https://docs.o-ran-sc.org/en/i-release/}},
}

@misc{ORAN-arch,
    title={{O-RAN Architecture Description 13.0 - O-RAN.WG1.TS.OAD-R004-v13.00}},
    author={{O-RAN Alliance}},
    year={2025}
}

@misc{ORAN-e2ap,
    title={{O-RAN E2 Application Protocol (E2AP) 8.0 - O-RAN.WG3.TS.E2AP-R004-v08.00}},
    author={{O-RAN Alliance}},
    year={2025}
}

@misc{ORAN-KPM,
    title={{O-RAN E2 Service Model (E2SM) KPM 6.0 - O-RAN.WG3.TS.E2SM-KPM-R004-v06.00}},
    author={{O-RAN Alliance}},
    year={2025}
}

@misc{ORAN-SEC,
    title={{O-RAN Security Threat Modeling and Risk Assessment 4.0 - O-RAN.WG11.Threat-Modeling.O-R004-v04.00}},
    author={{O-RAN Alliance}},
    year={2024}
}

@misc{ORAN-RICSEC,
    title={{O-RAN Study on Security for Near Real Time RIC and xApps - O-RAN.WG11.Security-Near-RT-RIC-xApps-TR.0-R003-v05.00}},
    author={{O-RAN Alliance}},
    year={2024}
}

@misc{ORAN-RC,
    title={{O-RAN E2 Service Model (E2SM), RAN Control 7.0 - O-RAN.WG3.TS.E2SM-RC-R004-v07.00}},
    author={{O-RAN Alliance}},
    year={2025}
}

@misc{3gpp-s3-254446-keyissue,
  author       = {{3GPP}},
  title        = {{S3-254446: Key Issue for MAC CE Protection}},
  howpublished = {3GPP SA3 contribution},
  year         = {2025},
  note         = {Security key issue discussion on MAC Control Element protection},
  url          = {https://www.3gpp.org/ftp/Meetings_3GPP_SYNC/SA3/docs/S3-254446.zip}
}

@misc{3gpp-s3-254157-keyissue,
  author       = {{3GPP}},
  title        = {{S3-254157: New key issue on MAC CE security }},
  howpublished = {3GPP SA3 contribution},
  year         = {2025},
  note         = {Security key issue discussion on MAC Control Element protection},
  url          = {https://www.3gpp.org/ftp/Meetings_3GPP_SYNC/SA3/docs/S3-254157.zip}
}

@misc{3gpp-s3-254352-keyissue,
  author       = {{3GPP}},
  title        = {{S3-254352: New Key issue on MAC CE security }},
  howpublished = {3GPP SA3 contribution},
  year         = {2025},
  note         = {Security key issue discussion on MAC Control Element protection},
  url          = {https://www.3gpp.org/ftp/Meetings_3GPP_SYNC/SA3/docs/S3-254352.zip}
}

@inproceedings{2021-SECURITY-CASideChannel,
  title={{A Stealthy Location Identification Attack Exploiting Carrier Aggregation in Cellular Networks}},
  author={Lakshmanan, Nitya and Budhdev, Nishant and Kang, Min Suk and Chan, Mun Choon and Han, Jun},
  booktitle={Proc. USENIX Security},
  year={2021}
}

@inproceedings{2020-WISEC-Botnet,
  title={Paging storm attacks against 4G/LTE networks from regional Android botnets: rationale, practicality, and implications},
  author={Fang, Kaiming and Yan, Guanhua},
  booktitle={Proc. ACM WiSec},
  pages={295--305},
  year={2020}
}

@inproceedings{2022-WinTECH-POWDER,
  title={NexRAN: Closed-loop RAN slicing in POWDER-A top-to-bottom open-source open-RAN use case},
  author={Johnson, David and Maas, Dustin and Van Der Merwe, Jacobus},
  booktitle={Proc. ACM MobiCom Workshop - WiNTECH},
  pages={17--23},
  year={2022}
}

@article{2023-comnet-ORANgym,
  title={OpenRAN Gym: AI/ML development, data collection, and testing for O-RAN on PAWR platforms},
  author={Bonati, Leonardo and Polese, Michele and D’Oro, Salvatore and Basagni, Stefano and Melodia, Tommaso},
  journal={Computer Networks},
  volume={220},
  pages={109502},
  year={2023},
  publisher={Elsevier}
}

@article{2013-commtut-ltesecsok,
  title={A survey on security aspects for LTE and LTE-A networks},
  author={Cao, Jin and Ma, Maode and Li, Hui and Zhang, Yueyu and Luo, Zhenxing},
  journal={IEEE communications surveys \& tutorials},
  volume={16},
  number={1},
  pages={283--302},
  year={2013},
  publisher={IEEE}
}

@inproceedings{2018-ISSCC-ltevulnsok,
  title={Security threats against LTE networks: A survey},
  author={Vachhani, Khyati},
  booktitle={Proc. SSCC},
  pages={242--256},
  year={2018},
  organization={Springer}
}

@inproceedings{2025-WISEC-5gvulnsok,
  title={SoK: Evaluating 5G-advanced protocols against legacy and emerging privacy and security attacks},
  author={Eleftherakis, Stavros and Giustiniano, Domenico and Kourtellis, Nicolas},
  booktitle={Proc. ACM WiSec},
  pages={196--210},
  year={2025}
}

@inproceedings{2019-NDSS-SESSIONCONF,
  title={Component-based formal analysis of 5G-AKA: Channel assumptions and session confusion},
  author={Cremers, Cas and Dehnel-Wild, Martin},
  booktitle={Proc. NDSS},
  year={2019},
  organization={Internet Society}
}

@inproceedings{2026-FutureG-TeleRAN,
  title={Towards Bridging the Telemetry Gap for Security Applications in 6G OpenRANs via eBPF},
  author={Wen, Haohuang and Yegneswaran, Vinod and Porras, Phillip and Gehani, Ashish and Sharma, Prakhar and Lin, Zhiqiang},
  booktitle={Proc. NDSS FutureG},
  year={2026}
}

@inproceedings{2022-USENIX-LTRACK,
  title={$\{$LTrack$\}$: Stealthy tracking of mobile phones in $\{$LTE$\}$},
  author={Kotuliak, Martin and Erni, Simon and Leu, Patrick and R{\"o}schlin, Marc and {\v{C}}apkun, Srdjan},
  booktitle={Proc. USENIX Security},
  pages={1291--1306},
  year={2022}
}

@inproceedings{2020-NDSS-IMP4GT,
  title={{IMP4GT: IMPersonation Attacks in 4G NeTworks}},
  author={Rupprecht, David and Kohls, Katharina and Holz, Thorsten and P{\"o}pper, Christina},
  booktitle={Proc. NDSS},
  volume={17},
  pages={26--60},
  year={2020}
}

@inproceedings{2020-USENIX-ReVoLTE,
  title={Call me maybe: Eavesdropping encrypted $\{$LTE$\}$ calls with $\{$ReVoLTE$\}$},
  author={Rupprecht, David and Kohls, Katharina and Holz, Thorsten and P{\"o}pper, Christina},
  booktitle={Proc. USENIX Security},
  pages={73--88},
  year={2020}
}

@inproceedings{2015-ISSTA-Plausibility,
    author = {Qi, Zichao and Long, Fan and Achour, Sara and Rinard, Martin},
    title = {{An Analysis of Patch Plausibility and Correctness for Generate-and-Validate Patch Generation Systems}},
    booktitle = {Proc. ACM ISSTA},
    year = {2015}
}

@article{2012-SE-Dynamictest,
  author  = {Christopher M. Hayden and Edward K. Smith and
             Eric A. Hardisty and Michael Hicks and Jeffrey S. Foster},
  title   = {Evaluating Dynamic Software Update Safety Using
             Systematic Testing},
  journal = {IEEE Transactions on Software Engineering},
  volume  = {38},
  number  = {6},
  pages   = {1340--1354},
  year    = {2012},
  doi     = {10.1109/TSE.2011.101}
}
}

\appendix
\section*{Ethical Considerations}
\label{appendix:ethics}
We strictly adhere to the ethics policy provided in our experiments.
No experiments were conducted on commercial cellular networks and only the test country codes are used for the MNC and MCC in our private test network, to ensure the experiments are conducted in an isolated and controlled environment.
To prevent interference with commercial networks, the attacks are only launched against our UEs, and the attacker devices never use credentials registered to a commercial SIM.

\section{Detailed Corpus Analysis}
\label{appendix:vulnerability-analysis}

\begin{figure}[t]
    \centering
    \includegraphics[width=\linewidth]{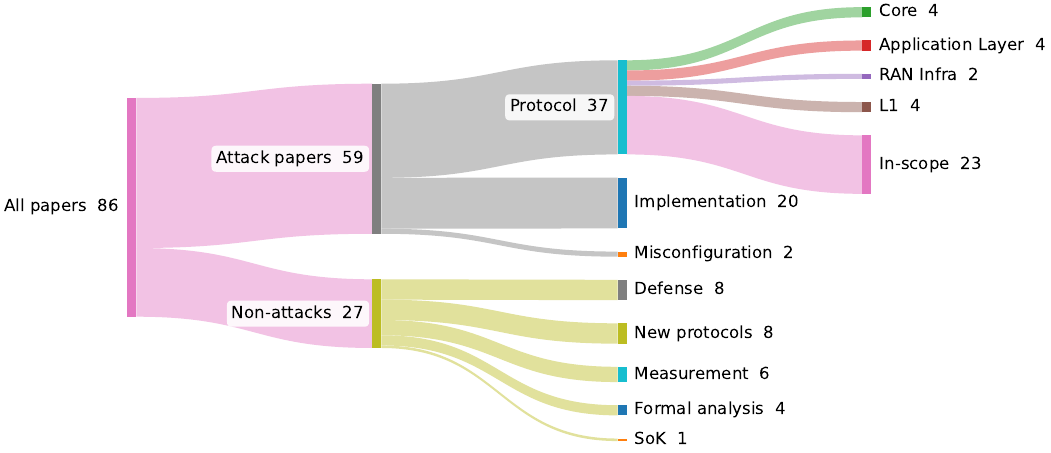}
    \vspace{-17pt}
    \caption{Breakdown of analyzed papers in our corpus.}
    \vspace{-13pt}
    \label{fig:corpus-sankey}
\end{figure}

The breakdown of all collected papers in our corpus is shown in Figure~\ref{fig:corpus-sankey}, the full list of analyzed attacks in Table~\ref{table:one-big-beautiful-table}, and the list of candidate hook points for hotfixes in Table~\ref{table:hooks}.
Below, we provide brief descriptions on the intuition and core idea of each hotfix not discussed in~\S\ref{subsec:hotfix-examples}.

\para{Installing null cipher/integrity~\cite{2019-CCS-5GReasoner}.}
Upon receiving \texttt{Security Mode Failure} during the RRC security context setup, limited service mode is activated for emergency calls, which uses null cipher and integrity algorithms.
To prevent adversaries from exploiting the limited service mode to extract private identifiers as described in the original paper, public 5G operators without regulatory requirements or private 5G operators who do not need this use case could explicitly release the UE upon security context setup failure. 

\para{SUCI-catcher~\cite{2021-WISEC-SUCI}.}
An MitM adversary sends a \texttt{Registration Request} with a reused SUCI value, which the RAN responds with a \texttt{Authentication Request}. 
Recently used SUCI values can be stored to conduct replay checks and reject requests with replayed values. 

\para{CAG list deletion~\cite{2024-USENIX-Hermes}.}
An MitM adversary forwards a victim UE's \texttt{Registration Request} to a RAN serving a network not included in the UE's allowed closed access group (CAG) list.
The RAN naturally responds with a \texttt{Registration Reject}, which causes the UE to delete its CAG list upon receipt, and does not reattempt connection, leading to DoS.
This attack is prevented by dropping \texttt{Registration Request} instead of sending an explicit response when it does not include the receiving RAN's serving CAG.
This allows the UE to timeout instead, and attempt reconnection shortly after.

\para{Uplink IMSI extractor~\cite{2022-MobiCom-Adaptover}.}
The \texttt{Attach Request} contains a random identity value in initial attach, or a previously-assigned temporary identifier (GUTI) for subsequent attaches.
The network requests for the permanent identifier (IMSI) in its \texttt{Identity Request} for initial attaches.
A signal injector adversary overshadows the attach request to contain a random identity value, leading to the UE revealing its IMSI.
A possible prevention to this is to modify the first \texttt{Identity Request} to always request the TMSI, a temporary identifier, and only fallback to a IMSI request if the UE fails to respond.
This prevents the UE from revealing its IMSI, while minimizing the impact for genuine first attach attempts.

\para{Remote de-registration~\cite{2019-SP-LTEFUZZ}.}
A malicious UE with a spoofed connection using another user's TMSI can send invalid NAS payloads, such as plaintext or replayed messages.
The NAS severs the connection to both the malicious and benign UE sharing the same TMSI via \texttt{Registration Reject}.
This attack is mitigated by preventing the spoofed connection in the first place similar to Blind DoS, or conducting sanity checks on the NAS payload and releasing the RRC connection for only the connection sending the invalid payload.

\para{SMS Phishing~\cite{2019-SP-LTEFUZZ}.}
A malicious UE with a spoofed connection using another user's TMSI sends an invalid (plaintext, replayed) SMS over NAS payload.
The NAS incorrectly accepts and processes the message allowing the adversary to send SMS messages impersonating as another user.
Spoofing prevention or sanity check on the payload can be conducted in a similar manner as the remote de-registration example.

\para{Selective service denial~\cite{2016-NDSS-Practical}.}
An MitM adversary downgrades the capability list in the UE's \texttt{Attach Request} to remove certain capabilities, selectively denying services.
Drawing inspiration from the possible mitigation discussed in the original paper, the RAN can store the capability list from the initial attach, and deny or restore downgraded capabilities.

\para{Authentication sync failure~\cite{2018-NDSS-LTEINSPEC}.}
A malicious UE repeatedly sends \texttt{Attach Request} spoofed as a victim UE to increase only the network's NAS counter, causing a de-sync between the victim UE and NAS.
As described in the original paper, the adversary is required to constantly change the security capability list for the network to recognize it as a new request and increment the counter.
Similar to the selective service denial example, the initially used security capability list for the UE can be stored and drop mismatching capabilities.

\para{Session confusion~\cite{2019-NDSS-SESSIONCONF}.}
A malicious UE sends two \texttt{Registration Request} messages in quick succession, first one using its own SUCI and the second using a victim UE's SUCI.
This causes a session confusion leading to the victim UE's anchor keys being compromised.
The attack can be mitigated by ensuring the \texttt{Registration Request} has been answered with \texttt{Authentication Request} first, dropping any additional \texttt{Registration Request} in between.

\para{Paging timing correlation~\cite{2019-NDSS-IMSICrack}.}
An adversary triggers paging toward a victim through a silent call or message, and correlates the timing of the resulting paging messages with its trigger to infer the victim's identity and presence in a cell.
The RAN populates each paging message with additional dummy identities, so that the identities observed following a trigger no longer isolate the victim.

\para{NAS counter reset~\cite{2019-CCS-5GReasoner}.}
An MitM adversary replays a previously captured \texttt{Security Mode Complete} message, which the network accepts because it carries a valid MAC, resetting the uplink NAS counter and allowing further replays under the reused counter value.
The RAN can compare the sequence number of an uplink \texttt{Security Mode Complete} against that of the \texttt{Security Mode Command} it transmitted, and drop a reply bearing zero to a command that did not, to keep the AMF's uplink counter aligned with the UE's.

\para{Keystream reuse~\cite{2020-USENIX-ReVoLTE}.}
The RAN reuses a dedicated bearer identity with identical key inputs for two successive calls within one RRC connection, producing the same keystream for both and allowing a passive adversary to recover the earlier call from the later one.
The RAN tracks the bearer identities already used on each connection, and rewrites the identity in an \texttt{RRC Connection Reconfiguration} that re-establishes a bearer with a previously released value to the next unused one.
Where every identity has been consumed within a connection, the RAN releases it instead, forcing fresh key derivation on reconnection.

\para{Authentication abort w/ attach, authentication abort w/ detach~\cite{2021-SP-Bookworm}.}
A malicious UE presents a victim's identity in an \texttt{Attach Request} or \texttt{Detach Request} and aborts before authentication completes, which the network treats as the victim relocating and uses to tear down the victim's existing session.
Similar to the Blind DoS, the RAN maintains the subscriber identity associated with each active connection, and drops an attach or detach carrying an identity already bound to a different connection that remains in the connected state.

\para{Radio resource draining~\cite{2021-Mobicom-IoTAttack}.}
A signal injector overshadows a victim's uplink transmission with a forged buffer status report claiming a large pending buffer, causing the RAN to commit uplink grants that starve other devices in the cell.
The RAN tracks how frequently large buffer sizes are reported on each connection, and drops the report once the rate exceeds what legitimate traffic produces.
The scheduler then never acts on the forged buffer, preserving service for the remaining devices in the cell.

\para{Uplink DoS~\cite{2022-MobiCom-Adaptover}.}
A signal injector overshadows a victim's uplink message so that the network replies with a reject carrying a persistent cause value, which incurs a long-term DoS.
By rewriting the cause in an outgoing \texttt{Attach Reject} to a non-persistent value, the UE retries shortly afterwards.
The initial rejection itself is unavoidable, but the long-term DoS impact is prevented, and the adversary must sustain the injection to keep the victim offline.

\para{Duplicate emergency attach~\cite{2022-Mobicom-Emergency}.}
Emergency attaches from anonymous UEs carry no security context, so an adversary can submit a duplicate \texttt{Attach Request} bearing a victim's IMEI, which the network cannot distinguish from a legitimate one and which implicitly detaches the victim's ongoing emergency session.
The RAN records the IMEI of each established emergency session, and drops a subsequent emergency attach presenting an IMEI already bound to an active session from a different connection.
The entry is cleared when the original connection is released, so a UE that genuinely reattaches after a reboot is delayed only until that release completes.

\section{On Detectability and Hotfixability}
\label{appendix:5gspector-hotfixability}
As discussed in \S\ref{subsec:corpus-results}, detectability and hotfixability capture related but different properties of an attack.
We provide further discussion on the differences between these two properties through a comparison with the detectable and hotfixable attacks in 5G-Spector's~\cite{2024-NDSS-5GSpector} attack catalogue.
Table~\ref{table:5gspector-hotfix} reproduces the catalogue as reported in 5G-Spector, and compares the 5G-Spector-detectability and \name-hotfixability coverages. 
While the two share a considerable overlap, some differences exist.

Of the 15 5G-Spector-detectable attacks, 7 are not \name-hotfixable, as \name-hotfixability requires a more restrictive detection requirement than 5G-Spector.
The 5G-Spector-detectable attacks contains both \emph{preventive detection}, for which detection occurs before the attack takes effect through observation of cause, and \emph{postmortem detection}, for which the attack is only detected after it takes effect through observation of consequence.
Of the two, \name-hotfixability requires \emph{preventive detection}, as the RAN must remove a necessary attack precondition before the vulnerable behavior takes effect, and \emph{postmortem detection} lacks this temporal property.

For example, the downlink DoS~\cite{2022-MobiCom-Adaptover} attack is detectable with 5G-Spector but not \name hotfixable.
The downlink DoS attack involves an attacker overshadowing the downlink \texttt{Authentication Request} during the attach process with an \texttt{Attach Reject} or \texttt{Service Reject} message to cause the UE to abort the attach process.
The 5G-Spector detection rule checks whether a UE responds to an \texttt{Authentication Request} with neither an \texttt{Authentication Response} nor an \texttt{Authentication Failure}. 
This absence of an appropriate response is the effect of the DoS caused by the overshadowed message already delivered to the UE, which is sufficient for detection, but leaves no remaining opportunity to act on it. 

On the other hand, 1 not 5G-Spector-detectable attack is \name-hotfixable.
The TORPEDO attack~\cite{2019-NDSS-IMSICrack}, is a passive attack where an attacker observes paging message timing to deduce a target UE's paging occasion.
Such passive attacks do not involve anomalous messages or event for a 5G-Spector to detect attack occurrence, as the attack relies on a side-channel always present in a vulnerable message type.
In the case of the TORPEDO attack, although the vulnerable paging messages are always observable from the RAN, there exists no trigger for 5G-Spector to detect if it is currently being exploited.
Whereas for \name-hotfixability, observation of these vulnerable message types is a sufficient detection criterion.
Prevention does not strictly require distinguishing an attack instance from benign traffic, if the exploitable property can be removed from every instance of the vulnerable message.

\begin{table*}[tbh]
\centering
\caption{L3 attacks catalogued by 5G-Spector~\cite{2024-NDSS-5GSpector}, annotated with \name hotfixability.}
\label{table:5gspector-hotfix}
\resizebox{\linewidth}{!}{%
\begin{tabular}{c l c c c c l c c}
\hline
\textbf{No.} & \textbf{Attack} & \textbf{Ref} & \textbf{Adversary} & \textbf{Layer} & \textbf{Implication} & \textbf{Exploited L3 Message} & \textbf{Detectable} & \textbf{Hotfixable} \\
\hline

1  & Authentication Sync Failure   & \cite{2018-NDSS-LTEINSPEC} & UE       & NAS & Availability    & AttachRequest              & $\bigcirc$   & $\bigcirc$ \\
2  & Traceability Attack           & \cite{2018-NDSS-LTEINSPEC} & FBS      & NAS & Privacy         & SecModeCommand             & $\triangle$  & $\triangle$ \\
3  & Numb Attack                   & \cite{2018-NDSS-LTEINSPEC} & FBS      & NAS & Availability    & AuthenticationReject       & $\triangle$  & $\triangle$ \\
4  & Paging Channel Hijacking      & \cite{2018-NDSS-LTEINSPEC} & FBS      & RRC & Availability    & Paging                     & $\triangle$  & $\triangle$ \\
5  & Stealthy Kicking-off Attack   & \cite{2018-NDSS-LTEINSPEC} & FBS      & RRC & Availability    & Paging                     & $\triangle$  & $\triangle$ \\
6  & Panic Attack                  & \cite{2018-NDSS-LTEINSPEC} & FBS      & RRC & Availability    & Paging                     & $\triangle$  & $\triangle$ \\
7  & Energy Depletion Attack       & \cite{2018-NDSS-LTEINSPEC} & FBS      & RRC & Availability    & Paging                     & $\triangle$  & $\triangle$ \\
8  & Linkability Attack            & \cite{2018-NDSS-LTEINSPEC} & FBS      & RRC & Privacy         & Paging                     & $\triangle$  & $\triangle$ \\
9  & Detach/Downgrade Attack       & \cite{2018-NDSS-LTEINSPEC} & FBS      & NAS & Availability    & DetachRequest              & $\triangle$  & $\triangle$ \\
\hline

10 & NAS Counter Reset             & \cite{2019-CCS-5GReasoner} & FBS/MitM & NAS & Availability    & SecModeCommand/Complete    & $\bigcirc$   & $\bigcirc$ \\
11 & Uplink NAS Counter Desync     & \cite{2019-CCS-5GReasoner} & FBS/MitM & NAS & Availability    & SecModeCommand             & $\bigcirc$   & $\times$ \\
12 & Exposing NAS Sequence Number  & \cite{2019-CCS-5GReasoner} & Passive  & NAS & Privacy         & SecModeCommand/Complete    & $\times$     & $\times$ \\
13 & Neutralizing TMSI Refreshment & \cite{2019-CCS-5GReasoner} & MitM     & NAS & Privacy         & ConfigUpdateCommand        & $\times$     & $\times$ \\
14 & Cutting off the Device        & \cite{2019-CCS-5GReasoner} & MitM     & NAS & Availability    & RegRequest/DeRegRequest    & $\times$     & $\times$ \\
15 & DoS with RRC Setup Request    & \cite{2019-CCS-5GReasoner} & UE       & RRC & Availability    & ConnectionRequest          & $\bigcirc$   & $\bigcirc$ \\
16 & Installing Null Cipher/Integrity & \cite{2019-CCS-5GReasoner} & MitM  & RRC & Confidentiality & SecurityModeComplete       & $\bigcirc$   & $\bigcirc$ \\
17 & Lullaby Attack                & \cite{2019-CCS-5GReasoner} & FBS/MitM & RRC & Availability    & RRCReconfiguration         & $\bigcirc$   & $\times$ \\
18 & Incarceration Attack          & \cite{2019-CCS-5GReasoner} & FBS/MitM & RRC & Availability    & RRCReject                  & $\bigcirc$   & $\times$ \\
19 & Exposing TMSI/Paging/RNTI     & \cite{2019-CCS-5GReasoner} & MitM     & RRC & Privacy         & RRCRelease                 & $\times$     & $\times$ \\
\hline

20 & BTS Resource Depletion        & \cite{2019-SP-LTEFUZZ}     & UE       & RRC & Availability    & ConnectionRequest          & $\bigcirc$   & $\bigcirc$ \\
21 & Blind DoS                     & \cite{2019-SP-LTEFUZZ}     & UE       & RRC & Availability    & ConnectionRequest          & $\bigcirc$   & $\bigcirc$ \\
22 & Remote De-registration        & \cite{2019-SP-LTEFUZZ}     & UE       & NAS & Availability    & AttachRequest              & $\bigcirc$   & $\bigcirc$ \\
23 & AKA Bypass Attack             & \cite{2019-SP-LTEFUZZ}     & Core     & RRC & Confidentiality & SecModeCommand/Complete    & $\bigcirc$   & $\times$ \\
\hline

24 & TORPEDO                       & \cite{2019-NDSS-IMSICrack} & Passive  & RRC & Privacy         & Paging                     & $\times$     & $\bigcirc$ \\
25 & PIERCER                       & \cite{2019-NDSS-IMSICrack} & FBS      & RRC & Privacy         & Paging                     & $\triangle$  & $\triangle$ \\
26 & IMSI Cracking                 & \cite{2019-NDSS-IMSICrack} & FBS      & RRC & Privacy         & Paging                     & $\triangle$  & $\triangle$ \\
\hline

27 & Location Leak Attacks in LTE  & \cite{2016-NDSS-Practical} & FBS      & RRC/NAS & Privacy      & Miscellaneous              & $\triangle$  & $\triangle$ \\
28 & DoS Attacks in LTE            & \cite{2016-NDSS-Practical} & FBS      & RRC/NAS & Availability & Miscellaneous              & $\triangle$  & $\triangle$ \\
\hline

29 & Downlink IMSI Extractor       & \cite{2022-USENIX-LTRACK}  & MitM     & NAS & Privacy         & IdentityRequest            & $\bigcirc$   & $\times$ \\
\hline

30 & Uplink IMSI Extractor         & \cite{2022-MobiCom-Adaptover} & MitM  & NAS & Privacy         & AttachRequest              & $\bigcirc$   & $\bigcirc$ \\
31 & Downlink DoS                  & \cite{2022-MobiCom-Adaptover} & MitM  & NAS & Availability    & AttachReject               & $\bigcirc$   & $\times$ \\
32 & Uplink DoS                    & \cite{2022-MobiCom-Adaptover} & MitM  & NAS & Availability    & AttachRequest              & $\bigcirc$   & $\bigcirc$ \\
\hline
\multicolumn{9}{c}{$\bigcirc$ = Detectable / hotfixable \qquad $\times$ = Not detectable / not hotfixable \qquad $\triangle$ = Tagged as addressed by FBS detection techniques in 5G-Spector} \\
\end{tabular}
}
\end{table*}

\begin{figure}[t]
    \centering
    \includegraphics[width=0.84\linewidth]{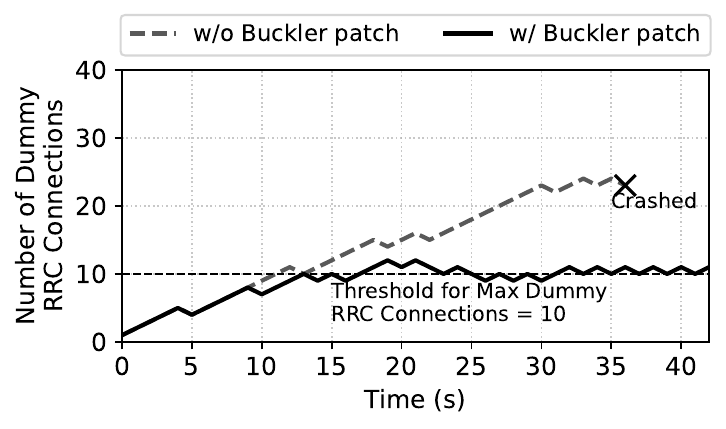}
    \vspace{-8pt}
    \caption{Evaluation of the BTS resource depletion hotfix on OAI RAN loaded with the identical hotfix descriptor as srsRAN.}
    \vspace{-10pt}
    \label{fig:oai-portability}
\end{figure}

\section{Additional Hotfix Portability Evaluation}
\label{appendix:portability-evaluation}

To further evaluate cross-implementation portability, we load the BTS resource-depletion hotfix in Listing~\ref{lst:btsdepletion} (Appendix~\ref{appendix:patch-template-example}) on OAI without changing the descriptor, its rules, or operator-level parameters from the srsRAN evaluation.
OAI and srsRAN differ in their internal functions and data structures, but their \name adapters expose the same standardized message boundaries and state to the hotfix.
Figure~\ref{fig:oai-portability} shows that the hotfix reproduces the same behavior on OAI: without \name, attacker-created dummy RRC connections continue to accumulate until the RAN crashes, whereas the hotfix bounds them around the configured threshold of ten connections.
This result confirms that the hotfix logic can be reused across the two vRAN implementations, with stack-specific adaptation confined to the hook adapters rather than the operator-provided descriptor.
\section{Patch Rule Descriptor Examples}
\label{appendix:patch-template-example}

\begin{lstlisting}[caption={Example rule descriptor for blind DoS}, label={lst:blinddos}]
codelets:
- channel: ulccch
  datastructs:
  - name: tmsi_rnti_map_out
    type: bpf_map
    map_type: hashmap
    key_type: uint32_t
    value_type: uint32_t
    max_entries: 8
  - name: rnti_tmsi_map_out
    type: bpf_map
    map_type: hashmap
    key_type: uint32_t
    value_type: uint32_t
    max_entries: 8

  handlers:
  - msg_type: rrcConnectionRequest
    actions:
    - action_type: extract_field
      nas: false
      name: tmsi
      check:
        field: criticalExtensions.choice.rrcConnectionRequest_r8.ue_Identity.present
        value: LTE_InitialUE_Identity_PR_s_TMSI
      field: criticalExtensions.choice.rrcConnectionRequest_r8.ue_Identity.choice.s_TMSI.m_TMSI
      type: uint32_t
      offset: 0
      then:
      - action_type: lookup
        target: tmsi_rnti_map_out
        name: old_rnti
        key: tmsi
      - action_type: if
        condition:
          cond_type: isnull
          op: old_rnti
        then:
        - action_type: update_map
          target: tmsi_rnti_map_out
          key: tmsi
          value: rnti
        - action_type: update_map
          target: rnti_tmsi_map_out
          key: rnti
          value: tmsi
        else:
        - action_type: if
          condition:
            op: '!='
            a: old_rnti
            b: rnti
          then:
          - action_type: drop

- channel: dldcch
  datastructs:
  - name: tmsi_rnti_map_in
    type: bpf_map
    map_type: hashmap
    key_type: uint32_t
    value_type: uint32_t
    max_entries: 8
  - name: rnti_tmsi_map_in
    type: bpf_map
    map_type: hashmap
    key_type: uint32_t
    value_type: uint32_t
    max_entries: 8

  handlers:
  - msg_type: rrcConnectionReconfiguration
    actions:
    - action_type: extract_field
      nas: true
      message_type: 66
      name: tmsi
      eid: 80
      type: uint32_t
      offset: 8
      then:
      - action_type: update_map
        target: tmsi_rnti_map_in
        key: tmsi
        value: rnti
      - action_type: update_map
        target: rnti_tmsi_map_in
        key: rnti
        value: tmsi
  - msg_type: rrcConnectionRelease
    decode: false
    actions:
    - action_type: lookup
      target: rnti_tmsi_map_in
      name: tmsi
      key: rnti
    - action_type: if
      condition:
        cond_type: notnull
        op: tmsi
      then:
      - action_type: delete_map
        target: tmsi_rnti_map_in
        key: tmsi
      - action_type: delete_map
        target: rnti_tmsi_map_in
        key: rnti

linked_datastructs:
- codelet_name: ulccch
  linked_codelet_name: dldcch
  map_name: tmsi_rnti_map_out
  linked_map_name: tmsi_rnti_map_in
- codelet_name: ulccch
  linked_codelet_name: dldcch
  map_name: rnti_tmsi_map_out
  linked_map_name: rnti_tmsi_map_in

\end{lstlisting}

\begin{lstlisting}[caption={Example rule descriptor for BTS depletion}, label={lst:btsdepletion}]
codelets:
- channel: dlccch
  datastructs:
    - name: rrc_conn_counter_in
      type: bpf_map
      map_type: array
      key_type: int
      value_type: uint32_t
      max_entries: 3
    - name: conn_rnti_list_in
      type: bpf_map
      map_type: array
      key_type: int
      value_type: uint32_t
      max_entries: 10

  handlers:
    - msg_type: rrcConnectionSetup
      decode: false
      actions:
        - action_type: set
          name: cnt_index
          type: uint32_t
          value: 0
        - action_type: set
          name: in_index
          type: uint32_t
          value: 1
        - action_type: set
          name: out_index
          type: uint32_t
          value: 2
        - action_type: lookup
          target: rrc_conn_counter_in
          name: cnt
          key: cnt_index
        - action_type: lookup
          target: rrc_conn_counter_in
          name: in
          key: in_index
        - action_type: lookup
          target: rrc_conn_counter_in
          name: out
          key: out_index

        - action_type: lookup
          target: conn_rnti_list_in
          name: release_rnti
          key: in
        - action_type: update_map
          target: conn_rnti_list_in
          key: in
          value: rnti
        - action_type: math
          assign: in
          expr:
            op: mod
            a:
              op: add
              a: in
              b: 1
            b: 10
        - action_type: update_map
          target: rrc_conn_counter_in
          key: in_index
          value: in
        - action_type: math
          assign: cnt
          expr:
            op: add
            a: cnt
            b: 1
        - action_type: update_map
          target: rrc_conn_counter_in
          key: cnt_index
          value: cnt

        - action_type: if
          condition:
            op: '>'
            a: cnt
            b: 10
          then:
            - action_type: math
              assign: out
              expr:
                op: mod
                a:
                  op: add
                  a: out
                  b: 1
                b: 10
            - action_type: update_map
              target: rrc_conn_counter_in
              key: out_index
              value: out
            - action_type: release
              id: release_rnti

- channel: dldcch
  datastructs:
    - name: rrc_conn_counter_out
      type: bpf_map
      map_type: array
      key_type: int
      value_type: uint32_t
      max_entries: 3

  handlers:
    - msg_type: rrcConnectionRelease
      decode: false
      actions:
        - action_type: set
          name: cnt_index
          type: uint32_t
          value: 0
        - action_type: lookup
          target: rrc_conn_counter_out
          name: cnt
          key: cnt_index
        - action_type: math
          assign: cnt
          expr:
            op: sub
            a: cnt
            b: 1
        - action_type: update_map
          target: rrc_conn_counter_out
          key: cnt_index
          value: cnt

linked_datastructs:
- codelet_name: dlccch
  linked_codelet_name: dldcch
  map_name: rrc_conn_counter_in
  linked_map_name: rrc_conn_counter_out
\end{lstlisting}

\begin{lstlisting} [caption={Example rule descriptor for CA side channel}, label={lst:caside}]
codelets:
  - channel: dlsch

    handlers:
    - msg_type: SCELL_ACTIVATION
      decode: false
      actions:
        - action_type: math
          assign: "*p"
          expr:
            op: or
            a: "*p"
            b:
              op: and
              a:
                random: true
              b: "0xFE"
\end{lstlisting}

\end{document}